\documentclass[fleqn,usenatbib]{mnras}

\usepackage{newtxtext,newtxmath}
\usepackage{float}
\usepackage[T1]{fontenc}
\usepackage{ae,aecompl}
\usepackage{graphicx}
\usepackage{tabularx}
\usepackage{amsmath}
\usepackage{amssymb}
\usepackage{scalerel}
\usepackage{natbib}
\usepackage{url}
\usepackage[flushleft]{threeparttable}
\usepackage{balance}
\newcommand{\nodata}{\centering\arraybackslash --}
\usepackage{microtype}

\title[AT\,2017fvz:\ a nova in NGC\,6822]{AT\,2017fvz:\ a nova in the dwarf irregular galaxy NGC\,6822}

\author[M. W. Healy et al.]{M. W. Healy,$^{1}$\thanks{E-mail: M.W.Healy@2017.ljmu.ac.uk}
M. J. Darnley,$^{1}$\thanks{E-mail: M.J.Darnley@ljmu.ac.uk}
C. M. Copperwheat,$^{1}$
A. V. Filippenko,$^{2,3}$\and
M. Henze,$^{4}$
J. C. Hestenes,$^{2}$
P. A. James,$^{1}$
K. L. Page,$^{5}$\and
S. C. Williams,$^{6}$
and W. Zheng$^{2}$ \\
$^{1}$Astrophysics Research Institute, Liverpool John Moores University, Liverpool, L3 5RF, UK\\
$^{2}$Department of Astronomy, University of California, Berkeley, CA 94720-3411, USA\\
$^{3}$Miller Senior Fellow, Miller Institute for Basic Research in Science, University of California, Berkeley, CA 94720, USA\\
$^{4}$Department of Astronomy, San Diego State University, San Diego, CA 92182, USA \\
$^{5}$X-Ray and Observational Astronomy Group, Department of Physics \& Astronomy, University of Leicester, LE1 7RH, UK\\
$^{6}$Physics Department, Lancaster University, Lancaster, LA1 4YB, UK\\
}

\date{Accepted 2019 April 15. Received 2019 April 15; in original form 2018 November 13}

\pubyear{2019}

\begin{document}
\label{firstpage}
\pagerange{\pageref{firstpage}--\pageref{lastpage}}
\maketitle

\begin{abstract}
A transient in the Local Group dwarf irregular galaxy NGC\,6822 (Barnard's Galaxy) was discovered on 2017 August 2 and is only the second classical nova discovered in that galaxy. We conducted optical, near-ultraviolet, and X-ray follow-up observations of the eruption, the results of which we present here. This `very fast' nova had a peak $V$-band magnitude in the range $-7.41>M_V>-8.33$\,mag, with decline times of $t_{2,V} = 8.1 \pm 0.2$\,d and $t_{3,V} = 15.2 \pm 0.3$\,d. The early- and late-time spectra are consistent with an \ion{Fe}{ii} spectral class. The H$\alpha$ emission line initially has a full width at half-maximum intensity of $\sim 2400$\,km\,s$^{-1}$ -- a moderately fast ejecta velocity for the class. The H$\alpha$ line then narrows monotonically to $\sim1800$\,km\,s$^{-1}$ by 70\,d post-eruption. The lack of a pre-eruption coincident source in archival \textit{Hubble Space Telescope} imaging implies that the donor is a main sequence, or possibly subgiant, star. The relatively low peak luminosity and rapid decline hint that AT\,2017fvz may be a `faint and fast' nova.
\end{abstract}


\section{Introduction}\label{Introduction}

Classical novae (CNe) belong to the class of accreting binaries known as cataclysmic variables. As first proposed by \citet{1954PASP...66..230W}, these are closely interacting binaries consisting of a white dwarf (WD) accreting material from a donor -- either a main sequence, subgiant, or red giant star \citep[see][]{2012ApJ...746...61D}. Through Roche-lobe overflow or the stellar wind of an evolved donor, hydrogen-rich material from the donor streams, usually via an accretion disk \citep{1995CAS....28.....W}, onto the WD where severe heating and compression take place. Given favourable conditions, this results in a thermonuclear runaway (TNR) within the accreted envelope on the WD with a proportion of that envelope subsequently ejected --- the nova eruption \citep{1976IAUS...73..155S}. The luminosity of these systems typically increases to a few $\times10^{4}\, L_{\odot}$ \citep[see, e.g.,][]{2010AN....331..160B} with peak absolute magnitudes reaching $M_V \approx -10.5$ in extreme cases \citep{2009ApJ...690.1148S,2018MNRAS.474.2679A}.

Following the TNR, stable H-burning continues within any material remaining on the WD surface. This results in the emission of a large amount of X-rays typically peaking in the range $30-50$\,eV --- the so-called super-soft X-ray source \citep[SSS; see][]{1992A&A...262...97V}. The SSS is initially obscured by  optically thick ejecta surrounding the nova; however, once the optical depth has decreased sufficiently, the SSS is unveiled.

All novae are predicted to recur \citep{2005ApJ...623..398Y}, but the broad range of times between consecutive eruptions has led to segregation based on recurrence period ($P_\mathrm{rec}$). CNe have been observed to erupt just once. Recurrent novae (RNe) are systems with a high-mass WD and high accretion rates that have been recorded erupting multiple times.

One can categorise novae into different speed classes based on their decline times, $t_2$ and $t_3$ \citep{1957gano.book.....G}. These denote the time taken to decay by 2 or 3 mag (respectively) from maximum light. \citet{1936PASP...48..191Z} first proposed a relationship between the decline time and the maximum absolute visual magnitude of a nova. Subsequently, \citet{1945PASP...57...69M} and \citet{1956AJ.....61...15A} developed the `maximum magnitude -- rate of decline' relation \citep[MMRD; see, e.g.,][]{2000AJ....120.2007D}. However, the MMRD suffers from a large scatter, and the relation has been diluted by the discovery of `faint and fast' novae \citep{2011ApJ...735...94K,2016ApJS..227....1S} and short-cycle RNe ($P_\mathrm{rec}<10$\,yr; Darnley \& Henze, in prep.).

Independent studies of Galactic novae using {\it Gaia} data release 2 \citep[DR2;][]{2018A&A...616A...1G} parallaxes appear to show contradictory results. \citet{2018MNRAS.481.3033S} proposes that the MMRD is an unusable distance determination method and that it should no longer be employed. However, \citet{2019A&A...622A.186S} show that the MMRD relationship is strengthened once {\it Gaia} distances are assumed. Although there is some overlap, these two studies use different samples of novae. \citet{2017ApJ...839..109S} used a large sample of M\,87 novae to clearly demonstrate (see their Figure~1) the impact of `faint and fast' novae on the MMRD distribution and therefore that the concept is inherently flawed.

Novae may be divided into two spectroscopic classes based on the prominent non-Balmer emission lines in their early-post-maximum spectra: either He/N or \ion{Fe}{ii} \citep{1992AJ....104..725W,1994ApJ...426..279W}. The contribution of novae from each class varies between different galaxies, possibly due to variations in the dominant stellar population and metallicity of a given host \citep{2013AJ....145..117S}. Novae from  younger (disk) populations have higher mean WD masses than those from older (bulge) populations. High-mass WDs create lower mass but higher velocity ejecta than their low-mass counterparts, and are believed to produce the He/N dominant spectra, with the lower mass WDs creating the \ion{Fe}{ii} class \citep{2012AJ....144...98W,2013AJ....145..117S}. 

The study of novae in extragalactic environments provides the only way to explore how the local environment (e.g., the metallicity and star-formation rate) can affect the nova rate and characteristics of nova eruptions \citep{2016ApJS..227....1S}. The M\,31 nova population is dominated by the \ion{Fe}{ii} class \citep[82\%;][]{2011ApJ...734...12S}. Yet `bulgeless' galaxies show similar numbers of each class. For example, five of the ten spectroscopically classified novae in M\,33 are \ion{Fe}{ii}; the M\,33 spectral type distribution differs from that of M\,31 at the 98\% confidence level \citep{2012ApJ...752..156S,2014ASPC..490...77S}. The fraction of \ion{Fe}{ii} novae in the LMC is also $\sim 50\%$ \citep{2013AJ....145..117S}. 

NGC\,6822 is a dark matter dominated \citep{2003MNRAS.340...12W} dwarf irregular galaxy in the Local Group at a distance $476\pm44$\,kpc \citep{2014ApJ...794..107R}. It provides a low-metallicity environment compared to the majority of Local Group novae: [Fe/H] $\approx -0.5$ \citep[see, e.g.,][]{2018A&A...613A..56L}.

AT\,2017fvz\footnote{\url{https://wis-tns.weizmann.ac.il/object/2017fvz}} (aka KAIT-17bm) is only the second nova to be discovered within NGC\,6822 (see Section~\ref{PreviousNova} for a discussion of the first nova). It was discovered on 2017 Aug.\ 2.384 UT with an unfiltered magnitude of 17.6 at $\alpha=19^\mathrm{h}45^\mathrm{m}1^\mathrm{s}\!.03$, $\delta=-14\degr46^\prime50^{\prime\prime}\!\!.74$ \citep*[J2000;][]{2017TNSTR.831....1H} by the Katzman Automatic Imaging Telescope \citep[KAIT;][]{2001ASPC..246..121F} of the Lick Observatory Supernova Search (LOSS). This nova was also observed by the All-Sky Automated Survey for Supernovae \citep[ASAS-SN; see][]{2014ApJ...788...48S} on Aug.\ 3.190 and then with the Asteroid Terrestrial-impact Last Alert System \citep[ATLAS;][]{2018PASP..130f4505T,2018AJ....156..241H} on Aug.\ 3.386.

Here we report optical, near-ultraviolet (NUV), and X-ray observations of the eruption of AT\,2017fvz. In Section~\ref{Observations and Data Analysis} we describe the observations and data analysis. In Section~\ref{Results} we present the results of the photometry, spectroscopy, and X-ray analysis, and we discuss these in Section~\ref{Discussion}. We summarise our findings in Section~\ref{Summary and Conclusions}. Throughout, all times are quoted in coordinated universal time (UT), all uncertainties are quoted to 1$\sigma$, and all upper limits to 3$\sigma$.

\section{Observations and Analysis}\label{Observations and Data Analysis}

\subsection{Ground-based photometry}
The field containing the nova had been monitored by KAIT using its clear filter since 2017 July 15.404 without any associated detections until the discovery on Aug.\ 2.384, after which the nova was followed until Aug.\ 31.284. ATLAS  monitored a similar field from July 5.477 using its `orange' filter, approximately covering $r'$ and $i'$ (5600--8200\,\AA)\footnote{\url{http://www.fallingstar.com/specifications.php}}, without any associated detections until the first detection on Aug.\ 3.386. Like KAIT, the nova was monitored after discovery by ATLAS for the next 47\,d until Sep.\ 19.317 using the orange filter and also a `cyan' filter which approximately covers $V$ and $r'$ (4200--6500\,\AA). A few hours before the ATLAS detection, the nova was detected by ASAS-SN on 2017 Aug.\ 3.190 with a $V$-band filter. A Liverpool Telescope \citep[LT;][]{2004SPIE.5489..679S} follow-up campaign began 7.53\,d post-discovery; observations were taken with IO:O\footnote{\url{http://telescope.livjm.ac.uk/TelInst/Inst/IOO}} through $u'BVr'i'$ filters. 

Debiasing and flatfielding of the LT data were performed by the automatic LT reduction pipeline. Aperture photometry was calculated from these data using standard tools within PyRAF and calibrated against stars from the Local Group Galaxies Survey \citep[LGGS;][]{2007AJ....133.2393M}. The $u'r'i'$ magnitudes of the LGGS stars were calculated using transformations from \citet[their Table~1]{2005AJ....130..873J}. Each time spectra were obtained with the SPectrograph for the Rapid Acquisition of Transients \citep[SPRAT;][see Section~\ref{Spectroscopy}]{2014SPIE.9147E..8HP} by the LT, acquisition images were also taken using the SPRAT detector. These  images were reduced in the same manner as the IO:O data. The acquisition images were unfiltered, but the photometry was calibrated relative to the $r'$ filter.

The KAIT data were reduced using a custom pipeline \citep{2010ApJS..190..418G}. Point-spread-function (PSF) photometry was obtained using the IDL implementation of DAOPHOT \citep{1987PASP...99..191S,1993ASPC...52..246L}. Several nearby stars from the APASS catalog \citep{2009AAS...21440702H} were used to calibrate the KAIT clear-band data, with their magnitudes converted to the Landolt $R$-band system using the empirical prescription presented by R.\ Lupton\footnote{\url{http://sdss.org/dr7/algorithms/sdssUBVRITransform.html}}.

ATLAS carries out difference imaging of every frame with respect to a reference sky and the photometry reported here is from those images. The photometry was carried out as described by \cite{2018PASP..130f4505T} and \cite{2017ApJ...850..149S}.

\subsection{Spectroscopy}\label{Spectroscopy}

The optical spectra of AT\,2017fvz were taken using SPRAT on the LT. SPRAT is a spectrograph with a slit $95^{\prime\prime}$ long and $1^{\prime\prime}\!\!.8$ wide giving a resolution of 18\,\AA\ per pixel, corresponding to $R\approx350$ at the centre of the spectrum. It covers visible wavelengths in the range 4000--8000\,\AA. The details of the spectra, which were obtained using the blue-optimised mode, are summarised in Table~\ref{SpectroscopyTable}. All spectra were extracted, wavelength calibrated, and flux calibrated using the SPRAT pipeline \citep{2014SPIE.9147E..8HP}, except for the Aug.\ 25 spectrum which was not flux calibrated owing to poor sky transparency (clouds). The spectra were then analysed using routines with PyRAF.

\begin{table}
\caption{LT SPRAT spectroscopy of AT\,2017fvz.}
\begin{center}
\label{SpectroscopyTable}
\begin{tabular}{lccc}
\hline
\hline
UT Date$^\mathrm{a}$ & MJD (d) & $t-t_0$ (d) & Exposure time (s) \\
\hline
\hline
2017-08-09.900 & 57974.900 & \phantom{0}8.016 & $3 \times 600$\\
2017-08-15.924 & 57980.924 & 14.040 & $3 \times 600$\\
2017-08-19.894 & 57984.894 & 18.010 & $3 \times 600$\\
2017-08-25.885 & 57990.885 & 24.001 & $3 \times 600$\\
2017-09-12.879 & 58008.879 & 41.995 & $3 \times 900$\\
2017-10-10.848 & 58036.848 & 69.964 & $3 \times 900$\\
\hline
\hline
\end{tabular}
\end{center}
\begin{tablenotes}
\small
\item $^\mathrm{a}$The date refers to the midpoint of each observation.
\end{tablenotes}
\end{table}

\subsection{\textit{Swift} NUV and X-ray observations}\label{UV and X-ray observations}

Five target-of-opportunity (ToO) observations with the Neil Gehrels {\it Swift} Observatory \citep{2004ApJ...611.1005G}, totalling 20.0\,ks, were utilised to follow the NUV and X-ray evolution of the AT\,2017fvz (Target ID:\ 10268). We summarise all of the \textit{Swift} data in Table~\ref{Swift}.

\begin{table*}
\caption{{\it Swift} UVOT photometry and XRT counts for AT\,2017fvz.}
\label{Swift}
\begin{center}
\begin{tabular}{ccccccccc}
\hline
\hline
Exp$^\mathrm{a}$ & Date$^\mathrm{b}$ & MJD & $t-t_0^\mathrm{c}$ & $uvw1^\mathrm{d}$ & \multicolumn{2}{c}{X-ray rate ($10^{-3}$\,ct\,s$^{-1}$)} & \multicolumn{2}{c}{$L^\mathrm{e}$ ($10^{37}$\,erg\,s$^{-1}$)} \\
(ks) & (UT) & (d) & (d) & (mag) & 0.3--1\,keV & 0.3--10\,keV & 0.3--1\,keV & 0.3--10\,keV \\
\hline
\hline
3.9 & 2017-09-09 & 58005 & \phantom{0}38.12 & $18.6\pm0.1$ & $<1.9$ & $<1.9$ & $<0.8$ & $<0.8$\\
3.7 & 2017-10-08 & 58034 & \phantom{0}67.12 & $19.4\pm0.2$ & $<3.5$ & $<3.3$ & $<1.4$ & $<1.3$\\
3.4 & 2017-11-07 & 58064 & \phantom{0}97.12 & $19.6\pm0.3$ & $<2.5$ & $<3.2$ & $<1.0$ & $<1.3$\\
3.7 & 2018-04-27 & 58235 & 268.12 & $20.2\pm0.3$ & $<3.4$ & $<3.2$ & $<1.3$ & $<1.3$\\
4.0 & 2018-08-25 & 58355 & 388.12 & $19.9\pm0.2$ & $<2.5$ & $<3.0$ & $<1.0$ & $<1.2$\\
\hline
\hline
\end{tabular}
\end{center}
\begin{tablenotes}
\small
\item $^\mathrm{a}$Dead-time corrected XRT exposure time.
\item $^\mathrm{b}$Start date of the observation.
\item $^\mathrm{c}$Time since day of eruption on 2017 Aug.\ 1.884.
\item $^\mathrm{d}$Vega magnitudes for the $uvw1$ filter (central wavelength:\ 2600\,\AA).
\item $^\mathrm{e}$X-ray luminosity upper limits (unabsorbed, blackbody fit, 0.3--1\,keV or 0.3--10\,keV, as indicated).
\end{tablenotes}
\end{table*}

NUV data were obtained with the UV/Optical Telescope \citep[UVOT;][]{2005SSRv..120...95R} through the $uvw1$ filter. X-ray data were collected by the X-ray Telescope \citep[XRT;][]{2005SSRv..120..165B} in photon-counting mode. The NUV data were processed with HEASoft tools \citep[v6.24;][]{1995ASPC...77..367B} and using the most recent calibration files. We extracted the count-rate upper limits from the X-ray data using the online {\it Swift} XRT tool\footnote{\url{http://www.swift.ac.uk/user_objects/}} \citep{2009MNRAS.397.1177E}.

\section{Results}\label{Results}

\subsection{Reddening}\label{Reddening}

NGC\,6822 has a Galactic longitude and latitude of $\ell= 25.4\degr$ and $b = -18.4\degr$, respectively \citep{1998ARA&A..36..435M}. This results in that galaxy being affected by a modest amount of foreground Milky Way extinction. \citet{1967AJ.....72..134K} found the Galactic reddening toward the outer regions of NGC\,6822 to be $E_{B-V}=0.27 \pm 0.03$\,mag, as did \citet{1995AJ....110.2715M} with $E_{B-V}=0.26$\,mag. These are consistent with \citet{1996AJ....112.1928G} and \citet{2007AJ....133.2393M} who found $E_{B-V}=0.24 \pm 0.03$\,mag and $E_{B-V}=0.25$\,mag, respectively. The online dust-mapping tool\footnote{\url{http://argonaut.skymaps.info}} \citep{2018MNRAS.478..651G} returns a Galactic reddening toward NGC\,6822 of $E_{B-V}=0.22 \pm 0.02$\,mag. 

Cepheid variables within NGC\,6822 have been employed to estimate the internal reddening. \citet{1983ApJ...273..539M} found $E_{B-V}=0.36$\,mag, \citet{2006ApJ...647.1056G} reported a similar average reddening of $E_{B-V}=0.36 \pm 0.01$\,mag. \citet{2014ApJ...794..107R} used optical and infrared data for Cepheids to determine that the foreground reddening along the line of sight to NGC\,6822 is $E_{B-V}=0.35\pm 0.04$\,mag.

We have no knowledge of the radial displacement of AT\,2017fvz within NGC\,6822 so we adopt the two most extreme values of reddening. The foreground reddening toward NGC\,6822 gives the lower limit, the addition of reddening internal to NGC\,6822 gives the upper limit.

\subsection{Photometric evolution}

\begin{figure*}
\centering
\includegraphics[width=0.94\textwidth]{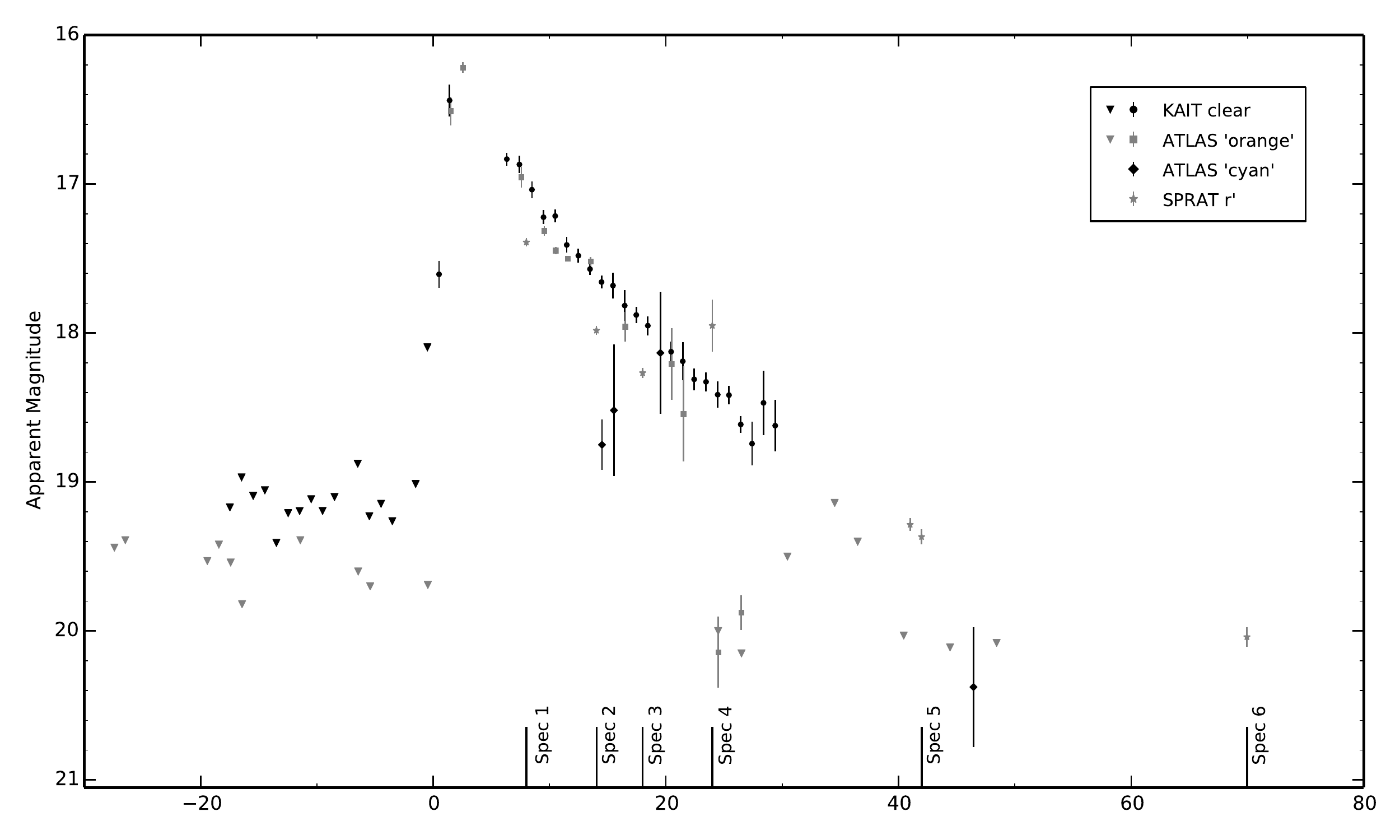}
\centering
\includegraphics[width=0.94\textwidth]{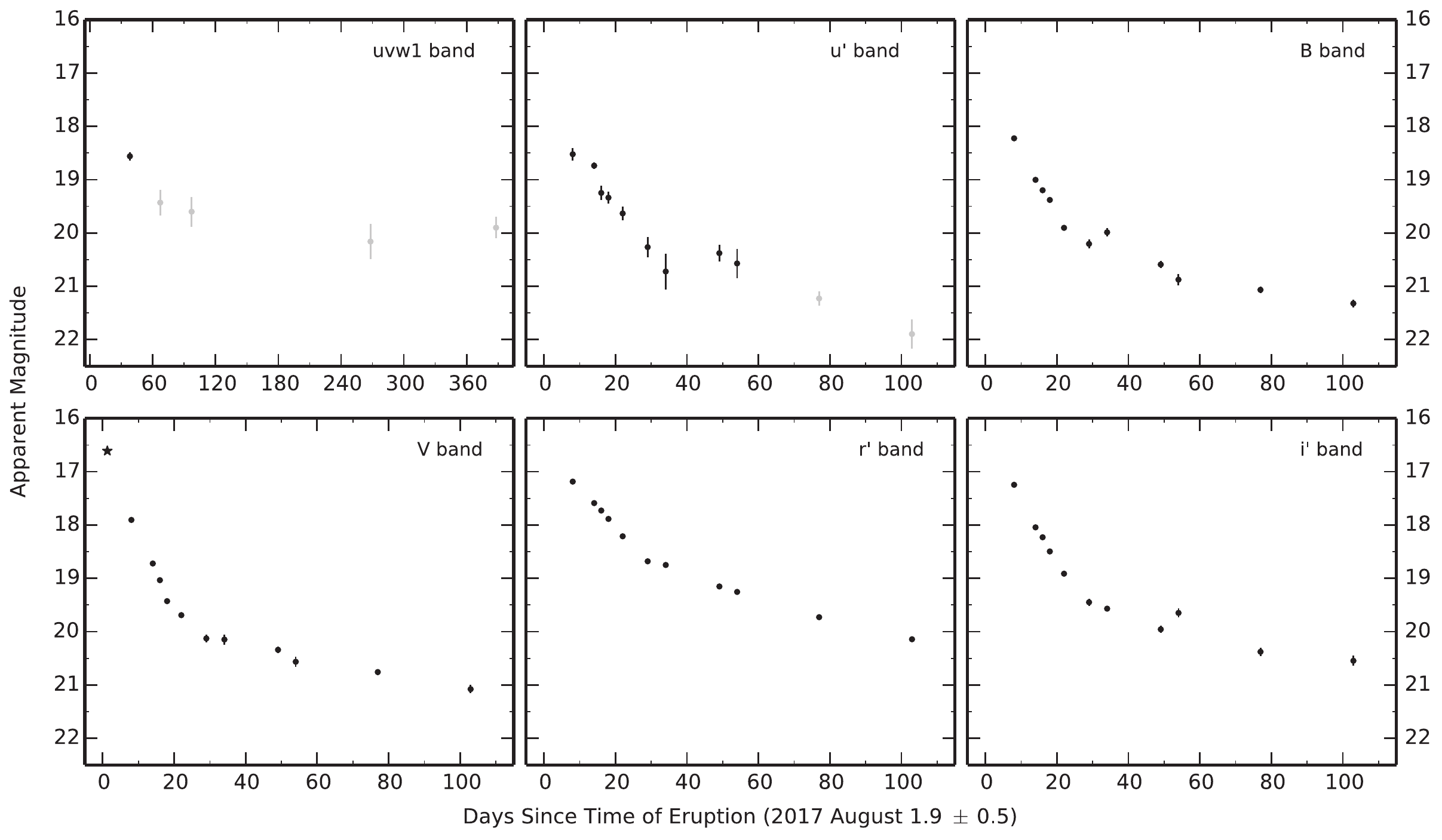}
\caption{Multiband light curves of AT\,2017fvz. Top:\ Optical light curve; see key for data sources. The epochs of the SPRAT spectra are indicated along the bottom axis. Lower panels:\ Light curves through individual filters. The star close to peak brightness in the $V$-band plot is from ASAS-SN. The grey points in the \textit{uvw1} and \textit{u'} panels are possibly contaminated by a nearby unresolved source (see Section~\ref{ProgenitorSection}). Note that the \textit{uvw1} plot covers $\sim420$\,d.}
\label{Lightcurve}\label{6Lightcurve}
\end{figure*}

The AT\,2017fvz photometry from ASAS-SN, ATLAS, KAIT, LT, and \textit{Swift} are presented in Figure~\ref{Lightcurve} and in Tables~\ref{Photometry1}--\ref{Photometry3}. The light curves illustrate that the nova was discovered prior to peak optical magnitude. We calculate the time of eruption to be 2017 Aug.\ $1.9 \pm 0.5$, the midpoint between the last nondetection (KAIT) with $m_\mathrm{clear} 
> 18.1$\,mag on 2017 Aug.\ 1.384 and the discovery on Aug.\ 2.384. Throughout, we refer to the time of eruption as $t_{0}$.

The $u'$, $B$, and $V$ bands all fade at approximately the same rate from peak until around 40\,d, while $i'$ fades more slowly and $r'$ even slower owing to the strong influence of the H$\alpha$ emission line on the broad-band photometry. We estimated the decline times ($t_2$ and $t_3$) of AT\,2017fvz by taking the brightest data point as the peak of the eruption and assuming a power-law decline (in luminosity) \citep[see, e.g.,][]{2006ApJS..167...59H}. The decline times for each filter are recorded in Table~\ref{Decline Times}. If we utilise the decline times with the MMRD relation \citep{2000AJ....120.2007D}, then we would expect peak absolute magnitudes of $M_{V} = -9.0 \pm 0.5$ and $M_{V} = -9.0 \pm 0.7$ for $t_2$ and $t_3$, respectively. It should be noted that this MMRD was derived from Galactic novae and, in addition to other limitations, may not be reliable within the differing environment of NGC\,6822 (see Section~\ref{PossibleRN}).

\begin{table}
\caption{Decline times of AT\,2017fvz in each filter.}
\label{Decline Times}
\begin{center}
\begin{tabular}{lrr}
\hline
\hline
Filter & $t_2$ (d) & $t_3$ (d) \\
\hline
\hline
$u'$ & $7.1 \pm 0.2$ & $13.7 \pm 0.3$ \\
$B$ & $7.0 \pm 0.2$ & $13.3 \pm 0.3$ \\
$V$ & $8.1 \pm 0.2$ & $15.2 \pm 0.3$ \\
$r'$ & $15.5 \pm 0.4$ & $33\phantom{.0} \pm 1\phantom{.0}$ \\
$i'$ & $13.0 \pm 0.3$ & $25.3 \pm 0.6$ \\
\hline
\hline
\end{tabular}
\end{center}
\end{table}

Taking the distance modulus of NGC\,6822 as $\mu_{0} = 23.38 \pm 0.02$\,mag \citep{2014ApJ...794..107R}, correcting for foreground reddening using $E_{B-V}=0.22 \pm 0.02$\,mag, we derive a lower limit for the peak absolute magnitude of $M_{V} = -7.41 \pm 0.07$. Here, we assumed that the peak observed magnitude (ASAS-SN) corresponded to the peak of the eruption. By extrapolating the $V$-band light curve power-law fit back to the final pre-eruption nondetection, we can estimate an upper limit on the peak eruption magnitude. Combining this upper limit with the estimate of the total (foreground and internal) reddening ($E_{B-V}=0.36 \pm 0.01$\,mag) yields an upper limit for the peak absolute magnitude of $M_{V} = -8.33 \pm 0.05$. 

There is evidence for a `plateau' in the $u'$, $B$, $V$, and $i'$-band light curves around $t = 25$\,d. As such, this nova would belong to the `plateau' class \citep[P-class;][]{2010AJ....140...34S}, where an otherwise smoothly declining light curve is interrupted by a short period when the optical magnitude remains approximately constant. 

\subsection{Spectroscopic evolution}

\begin{figure*}
\includegraphics[width=0.97\textwidth]{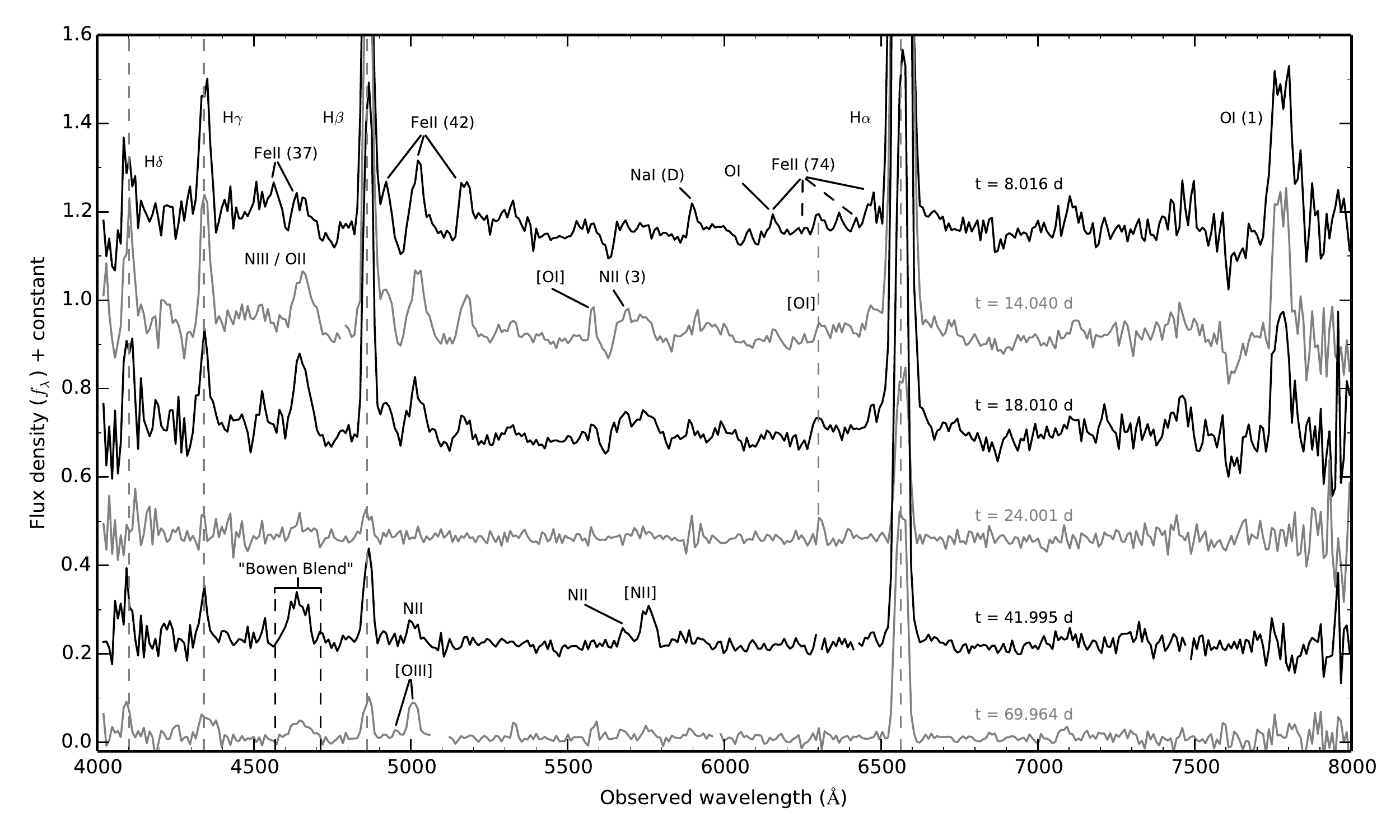}
\caption{The six SPRAT spectra of AT\,2017fvz, where the times post-eruption for each spectrum and prominent emission lines are indicated with labels. The gaps in the spectra at $t = 14.040$\,d, $t = 41.995$\,d, and $t = 69.964$\,d were the locations of substantial cosmic rays.}
\label{AllSpectra}
\end{figure*}

To aid the analysis of the AT\,2017fvz spectra, we made extensive use of spectral line data from \citet{1945CoPri..21....1M} and \citet{2012AJ....144...98W}. All of the spectra of AT\,2017fvz are plotted in Figure~\ref{AllSpectra}. These spectra can be split into three groups:\ the first contains the first four spectra that are within a $\sim16$-day-long period during early decline; the fifth spectrum was taken at $t \approx 42$\,d during the plateau, and the sixth was taken at $t \approx 70$\,d during the nebular phase. The spectra are presented in the rest frame of the observer. The average radial velocity of NGC\,6822 is $-57$\,km\,s$^{-1}$ \citep{2004AJ....128...16K}. The flux and full width at half-maximum intensity (FWHM) velocity were calculated by fitting Gaussian profiles to the emission lines using the SPLAT package in STARLINK; the fluxes and velocities are reported in Tables~\ref{SpecFlux} and \ref{SpecVel}, respectively.

\begin{table*}
\caption{The evolution of emission-line fluxes from the spectra of AT\,2017fvz in units of $10^{-15}$\,erg\,cm$^{-2}$\,s$^{-1}$.}
\label{SpecFlux}
\begin{center}
\begin{tabular}{lrrrrr}
\hline
\hline
Line identification & & & & & \\
(rest wavelength [\AA]) & $t = 8.016$\,d & $t = 14.040$\,d & $t = 18.010$\,d & $t = 41.995$\,d & $t = 69.964$\,d \\
\hline
\hline
H$\delta$ (4102) & 6 $\pm$ 3 & 8 $\pm$ 2 & 7 $\pm$ 3 & 2 $\pm$ 1 & 1.7 $\pm$ 0.4 \\
H$\gamma$ (4341) & 12 $\pm$ 1 & 9 $\pm$ 1 & 7 $\pm$ 1 & 2.4 $\pm$ 0.5 & 2.0 $\pm$ 0.6 \\
H$\beta$ (4861) & 30 $\pm$ 3 & 31 $\pm$ 4 & 27 $\pm$ 2 & 6.8 $\pm$ 0.5 & 2.4 $\pm$ 0.5 \\
{[\ion{O}{iii}]} (5007) & \nodata & \nodata & \nodata & \nodata & 2.6 $\pm$ 0.4 \\
\ion{Fe}{ii} (5018) & 4.6 $\pm$ 0.7 & 5 $\pm$ 1 & \nodata & \nodata & \nodata \\
H$\alpha$ (6563) & 139 $\pm$ 6 & 210 $\pm$ 10 & 230 $\pm$ 10 & 61 $\pm$ 3 & 24 $\pm$ 2\\
\ion{O}{i} (7773) & 28 $\pm$ 4 & 22 $\pm$ 3 & 16 $\pm$ 2 & \nodata & \nodata \\
\hline
\hline
\end{tabular}
\end{center}
\begin{tablenotes}
\small
\item We have not included the spectrum from 24.001\,d after eruption because the fluxes are not reliable.
\end{tablenotes}
\end{table*}

\begin{table*}
\caption{The evolution of emission-line velocities from the spectra of AT\,2017fvz in units of km\,s$^{-1}$.}
\label{SpecVel}
\begin{center}
\begin{tabular}{lrrrrrr}
\hline
\hline
Line identification & & & & & & \\
(rest wavelength [\AA]) & $t = 8.016$\,d & $t = 14.040$\,d & $t = 18.010$\,d & $t = 24.001$\,d & $t = 41.995$\,d & $t = 69.964$\,d \\
\hline
\hline
H$\delta$ (4102) & 2600 $\pm$ 900 & 2100 $\pm$ 300 & 2100 $\pm$ 600 & \nodata & 900 $\pm$ 50 & 1500 $\pm$ 300 \\
H$\gamma$ (4341) & 2500 $\pm$ 200 & 2000 $\pm$ 200 & 2200 $\pm$ 300 & \nodata & 1400 $\pm$ 200 & {\it 3500} $\pm$ {\it 800}$\ ^\mathrm{a}$ \\
H$\beta$ (4861) & 2300 $\pm$ 200 & 2100 $\pm$ 200 & 1900 $\pm$ 100 & \nodata & 1800 $\pm$ 100 & 1600 $\pm$ 200 \\
{[\ion{O}{iii}]} (5007) & \nodata & \nodata & \nodata & \nodata & \nodata & 1900 $\pm$ 200 \\
\ion{Fe}{ii} (5018) & 1900 $\pm$ 200 & 2200 $\pm$ 200 & \nodata & \nodata & \nodata & \nodata \\
H$\alpha$ (6563) & 2430 $\pm$ 70 & 2300 $\pm$ 100 & 2070 $\pm$ 70 & 2000 $\pm$ 90 & 1840 $\pm$ 60 & 1900 $\pm$ 100 \\
\ion{O}{i} (7773) & 2800 $\pm$ 300 & 2200 $\pm$ 200 & 2000 $\pm$ 200 & \nodata & \nodata & \nodata \\
\hline
\hline
\end{tabular}
\end{center}
\begin{tablenotes}
\small
\item $^\mathrm{a}$ This velocity is an upper limit as the H$\gamma$ line is blended with other lines around this wavelength.
\end{tablenotes}
\end{table*}

\subsubsection{Early Decline}\label{Early Decline}

\begin{figure*}
\includegraphics[width=0.95\textwidth]{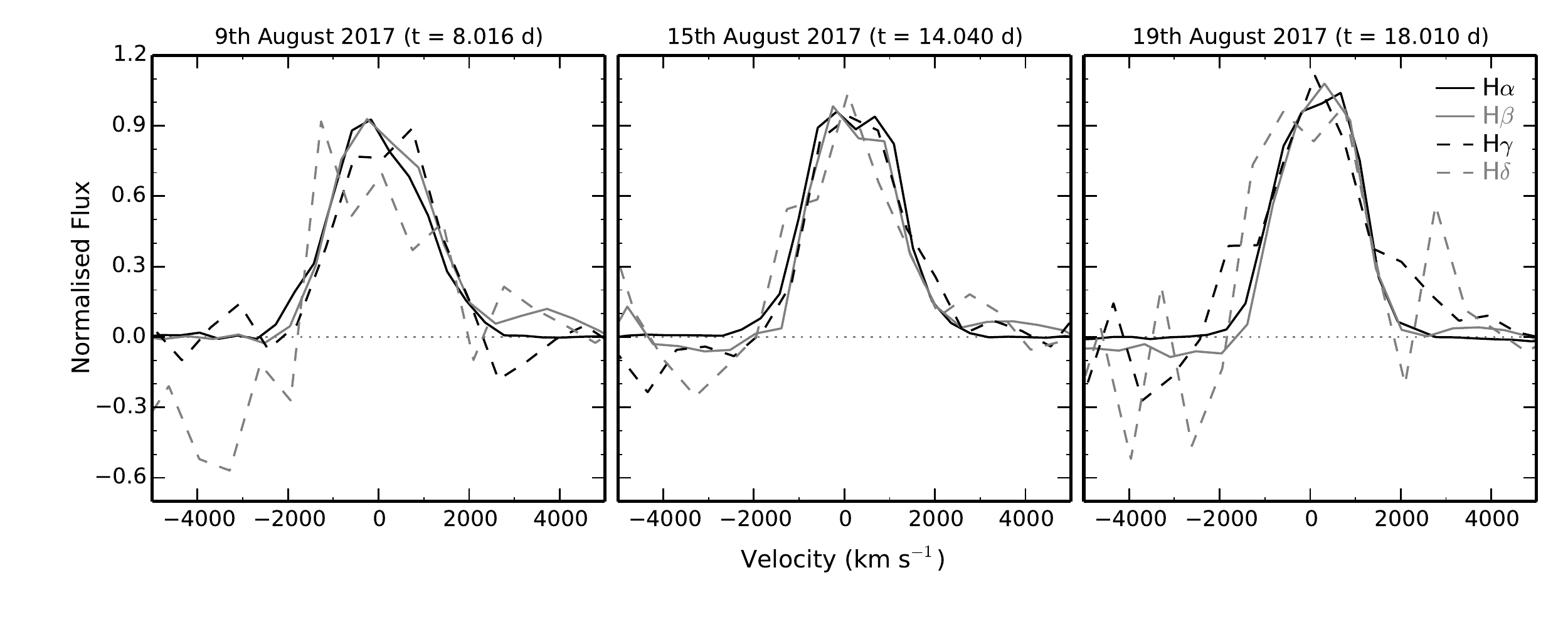}
\caption{The evolution of the AT\,2017fvz H$\alpha$--H$\delta$ emission lines over $\sim18$\,d post-eruption (normalised to peak flux). The H$\delta$ profile shows tentative evidence for blueshifted absorption ($t = 8.016$\,d) -- a possible P Cygni profile.}
\label{PCygni}
\end{figure*}

The first spectrum was taken at $t = 8.016$\,d when the nova was in the early decline phase \citep{2017ATel10630....1W}. By this time, we have missed the optically thick `fireball' stage which occurs on the rise until around peak brightness. We may have caught the very end of this transition with some of the \ion{Fe}{ii} lines and the H$\delta$ emission line still showing tentative signs of P\,Cygni profiles. The H$\delta$ line has a small blueshifted absorption component with a midpoint of $4052 \pm 8$\,\AA\ and an equivalent width of $29 \pm 6$\,\AA. Also, the emission component may have a different profile than the other Balmer lines with a FWHM of $\sim 2400$\,km\,s$^{-1}$. In Figure~\ref{PCygni} we show the Balmer lines from the first three spectra to illustrate the tentative evidence for an H$\delta$ P Cygni profile.

As the predominant non-Balmer emission lines are those of iron, AT\,2017fvz is consistent with the \ion{Fe}{ii} class. The broad Balmer lines lie close to the border value of $2500$\,km\,s$^{-1}$ which defines the broad-lined \ion{Fe}{ii} novae class \citep[\ion{Fe}{ii}b;][]{2009ApJ...690.1148S}.

The other prominent features of this first spectrum are the H$\alpha$, H$\beta$, and H$\gamma$ emission lines and the double-peaked \ion{O}{i} (1) emission line at 7773\,\AA, all of which have FWHM velocities of $\sim2400$\,km\,s$^{-1}$. There is an \ion{Fe}{ii} (42) triplet redward of H$\beta$ at 4924, 5018, and 5169\,\AA, as well as a fairly strong \ion{Na}{i}\,D emission line at $\sim 5892$\,\AA. There may be a weak \ion{Fe}{ii} (74) multiplet blueward of H$\alpha$ with 6148\,\AA\ and 6456\,\AA\ lines visible. However, with the 6248\,\AA\ and 6417\,\AA\ lines clearly absent, the feature at 6456\,\AA\ is more likely to be associated with nitrogen. Another explanation for this line at 6148\,\AA\ could be \ion{O}{i} $\lambda$6158. Between these lines, there is a feature at around 6300\,\AA, which is almost certainly [\ion{O}{i}], that persists until the `plateau' phase. We also see tentative evidence for \ion{Fe}{ii} (37) lines at 4556\,\AA\ and 4629\,\AA.

The second AT\,2017fvz spectrum, taken 14.040\,d post-eruption, maintains all aforementioned emission lines including the \ion{H}{i} lines, which show slightly lower FWHM velocities of $\sim 2100$\,km\,s$^{-1}$, and many \ion{Fe}{ii} lines. In addition, a prominent feature has developed at around 4640\,\AA\ which may be a blend of \ion{N}{iii} and \ion{O}{ii} emission lines at 4638\,\AA\ and 4676\,\AA, respectively (see Section~\ref{`Plateau' Phase}). Other emission lines could be present at this location including \ion{C}{iv} $\lambda$4658, [\ion{Fe}{iii}] $\lambda$4658, or \ion{O}{i} (18) at 4655\,\AA. The [\ion{O}{i}] $\lambda$5577 line can be seen alongside the \ion{N}{ii} (3) line at 5679\,\AA, similar to V1494\,Aquil\ae\ (Nova Aql 1999) $\sim14$\,d post-maximum \citep[see their Figure 6]{2003A&A...404..997I} and to SN\,2010U\footnote{Not a supernova \citep{2013ApJ...765...57C}!}, 15.3\,d post-maximum \citep[see their Figure 11]{2013ApJ...765...57C}.

The third spectrum ($t = 18.010$\,d) is similar to the previous two, with all lines except H$\alpha$ (see Section~\ref{HaEvolution}), H$\beta$, and the blend at $\sim 4640$\,\AA\ having weakened. Unfortunately, the fourth spectrum ($t=24.001$\,d) has low signal-to-noise ratio owing to poor observing conditions, and only Balmer and [\ion{O}{i}] $\lambda$6300 lines are apparent.

\subsubsection{`Plateau' Phase}\label{`Plateau' Phase}
The fifth spectrum was taken 41.995\,d post-eruption during the apparent plateau phase. The \ion{H}{I} emission lines still dominate, but these are joined by nitrogen emission lines such as \ion{N}{ii} (24) at 5001\,\AA, \ion{N}{ii} (3) at 5679\,\AA, and [\ion{N}{ii}] $\lambda$5755. During this evolutionary phase, we might expect to see a considerable enhancement of nitrogen lines --- the so-called `nitrogen flaring' --- caused by the Bowen fluorescence mechanism whereby \ion{N}{iii} is `pumped' by the UV resonance lines of \ion{O}{iii} \citep{1934PASP...46..146B,1935ApJ....81....1B}. \citet{2018A&A...611A...3H} suggested that this `Bowen Blend' ($\sim4640$\,\AA) may be more naturally explained by `oxygen flaring,' whereby there is flaring of the \ion{O}{ii} multiplet (V1) in the range 4638--4696\,\AA.

Such N- or O-flaring may manifest in the spectrum of AT\,2017fvz through a broad amalgamation of lines at approximately 4640\,\AA, where it is difficult to distinguish the individual lines owing to the low spectral resolution. We assume that they are the \ion{N}{ii} (5) multiplet at 4614\,\AA, 4621\,\AA, and 4630\,\AA, and the \ion{C}{iii} (1) multiplet at 4647\,\AA, 4650\,\AA, and 4651\,\AA, as well as other nitrogen species.

\subsubsection{Nebular Phase}
The final spectrum was taken 69.964\,d post-eruption. Here, there is evidence for the [\ion{O}{iii}] nebular lines at 4959 and 5007\,\AA. Only a handful of novae beyond the Magellanic Clouds have been observed spectroscopically during this phase \citep{Williams_2017}. The appearance of [\ion{O}{iii}] often roughly coincides with the beginning of the SSS phase when the ejecta from the nova are becoming optically thin to UV radiation and collisions are still occurring owing to the sufficiently high density providing a cooling mechanism \citep{2018ApJ...853...27M}. Additionally, the `Bowen Blend' is still visible but has broadened and taken on a `dome-like' appearance. 

At this time, the density of the ejecta must be less than the critical density ($n_e^\mathrm{crit} = 6.8 \times 10^5$\,cm$^{-3}$) for the collisional de-excitation of [\ion{O}{iii}]. One might also expect the [\ion{O}{iii}] auroral line at 4363\,\AA; however, as this is a relatively weak line, it is most likely blended with H$\gamma$ or hidden by the \ion{Hg}{i} night-sky line at 4358\,\AA. Even so, we can use the ratio of these three emission lines to estimate an upper limit for the electron temperature within this part of the ejecta of 5000\,K \citep[see Figure 5.1 in][]{2006agna.book.....O}.

\subsubsection{H$\alpha$ evolution}\label{HaEvolution}

\begin{figure}
\centering
\includegraphics[width=0.95\columnwidth]{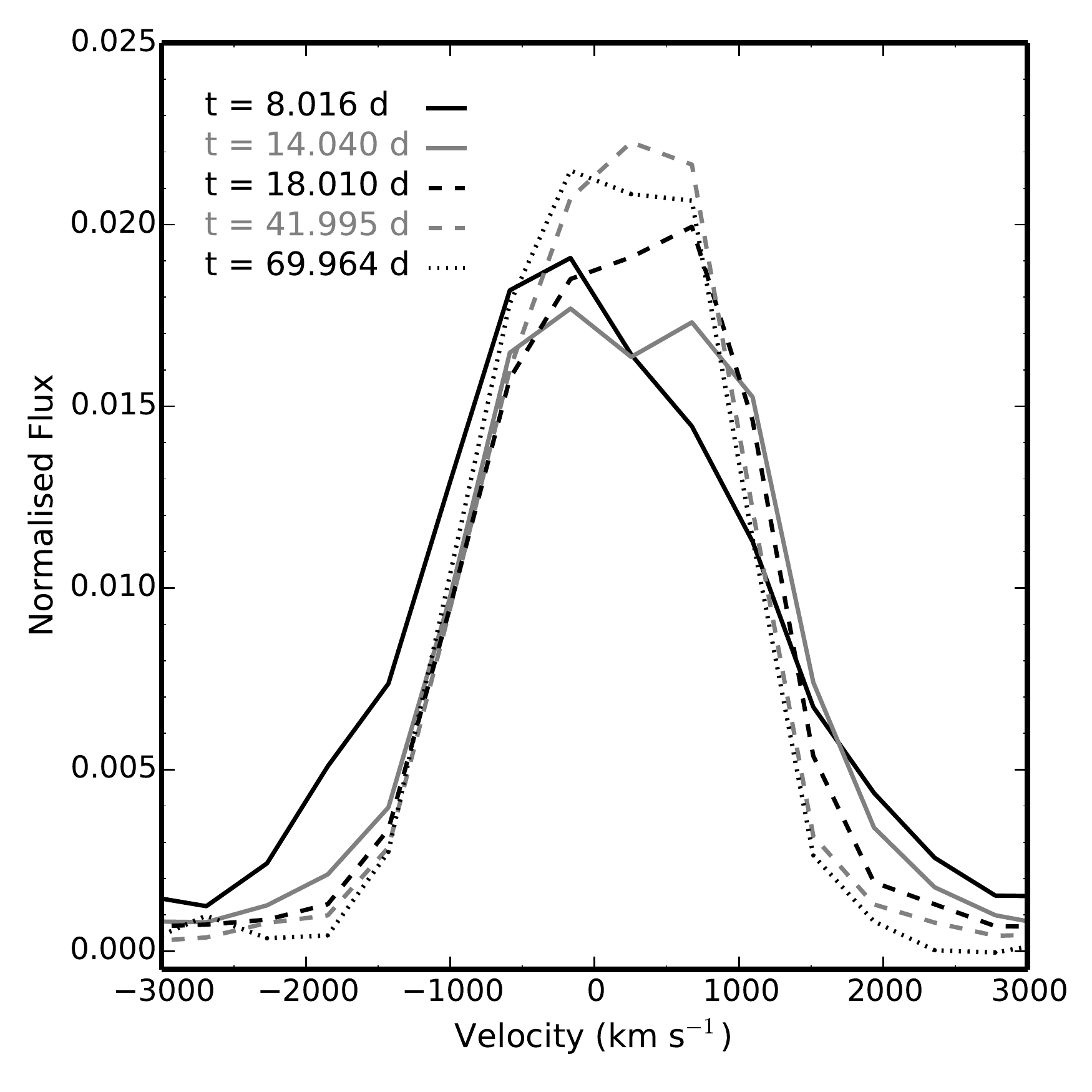}
\caption{H$\alpha$ emission-line evolution from $t = 8$\,d to $t = 70$\,d in terms of normalised flux and velocity. The $t = 24$\,d spectrum is not included owing to lack of flux calibration.}
\label{AllVelocityHaBoth}
\end{figure}

The evolution of the H$\alpha$ emission-line profile is shown in Figure~\ref{AllVelocityHaBoth}. After the first spectrum at $t \approx 8$\,d, when the line has a FWHM of $2430 \pm 70$\,km\,s$^{-1}$, the line progressively narrows from $2300 \pm 100$\,km\,s$^{-1}$ to $2070 \pm 70$\,km\,s$^{-1}$ and then $2000 \pm 90$\,km\,s$^{-1}$ at $t \approx 14$\,d, $t \approx 18$\,d, and $t \approx 24$\,d, respectively. The line width then remains constant between the fifth and sixth spectra with the FWHM being $1840 \pm 60$\,km\,s$^{-1}$ at $t \approx 42$\,d and $1900 \pm 100$\,km\,s$^{-1}$ at $t \approx 70$\,d. With no evidence for substantial circumbinary material, such line narrowing is probably due to decreasing emissivity as the ejecta expand, rather than a deceleration.

\subsection{X-rays}\label{X-rays}
Utilising the $r'$-band decline time ($t_{2} \approx 15$\,d; see Table~\ref{Decline Times}), we used the correlations presented by \citet{2014A&A...563A...2H} to predict the expected SSS properties of AT\,2017fvz. These indicate that a SSS with blackbody temperature $kT \approx 50$\,eV should have appeared at $t_\mathrm{on}\approx72$\,d and turned off at $t_\mathrm{off}\approx243$\,d. 

A Galactic foreground column density of $N_\mathrm{H} = 10^{21}$\,cm$^{-2}$ toward AT\,2017fvz was derived from the HEASARC $N_\mathrm{H}$ tool based on the Galactic neutral hydrogen map by \citet{2005A&A...440..775K}. We used the PIMMS software (v4.8f) with this column and the estimated SSS temperature to convert from counts to unabsorbed flux. We then derived X-ray luminosities by assuming a distance of 476\,kpc to NGC\,6822; these are presented in Table~\ref{Swift}.

We do not detect any X-ray emission from AT\,2017fvz in any of the five visits between 38\,d and 388\,d post-eruption. The luminosity upper limits, calculated from the X-ray count limits (0.3--1\,keV), assuming $kT \approx 50$\,eV in Table~\ref{Swift}, are all below $1.4 \times 10^{37}$\,erg\,s$^{-1}$. The assumed temperature is low compared in particular to fast RNe such as M31N\,2008-12a \citep[$\sim 120$\,eV;][]{2016ApJ...833..149D} and RS\,Oph \citep[$\sim 90$\,eV;][]{2011ApJ...727..124O}; therefore,  AT\,2017fvz must not have had a bright SSS phase during our observational window.

\subsection{The nova progenitor}\label{ProgenitorSection}
A nova system may harbour either a main sequence, subgiant, or red giant donor. If AT\,2017fvz hosted a red giant or a luminous accretion disk then it could have been detectable with \textit{Hubble Space Telescope} (\textit{HST}) owing to the proximity of NGC\,6822 \citep{2014ApJS..213...10W}. AT\,2017fvz is located within archival \textit{HST} Wide-Field Planetary Camera 2 (WFPC2) images (GO-11079) taken through the F170W, F255W, F336W, F439W, F555W, and F814W filters.

\begin{figure*}
\centering
\includegraphics[width=0.323333\textwidth]{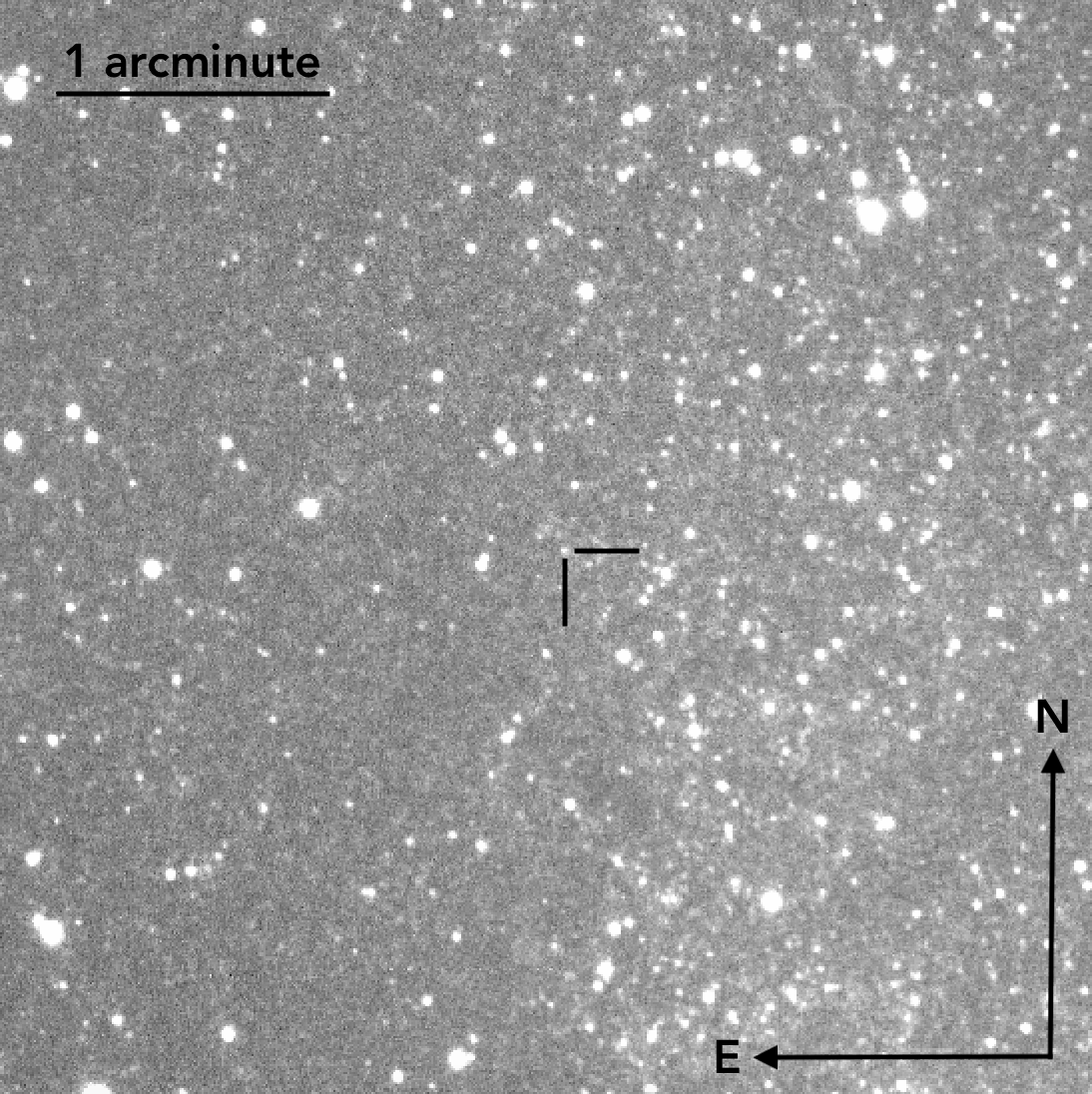}
\label{NovaLT}
\centering
\includegraphics[width=0.323333\textwidth]{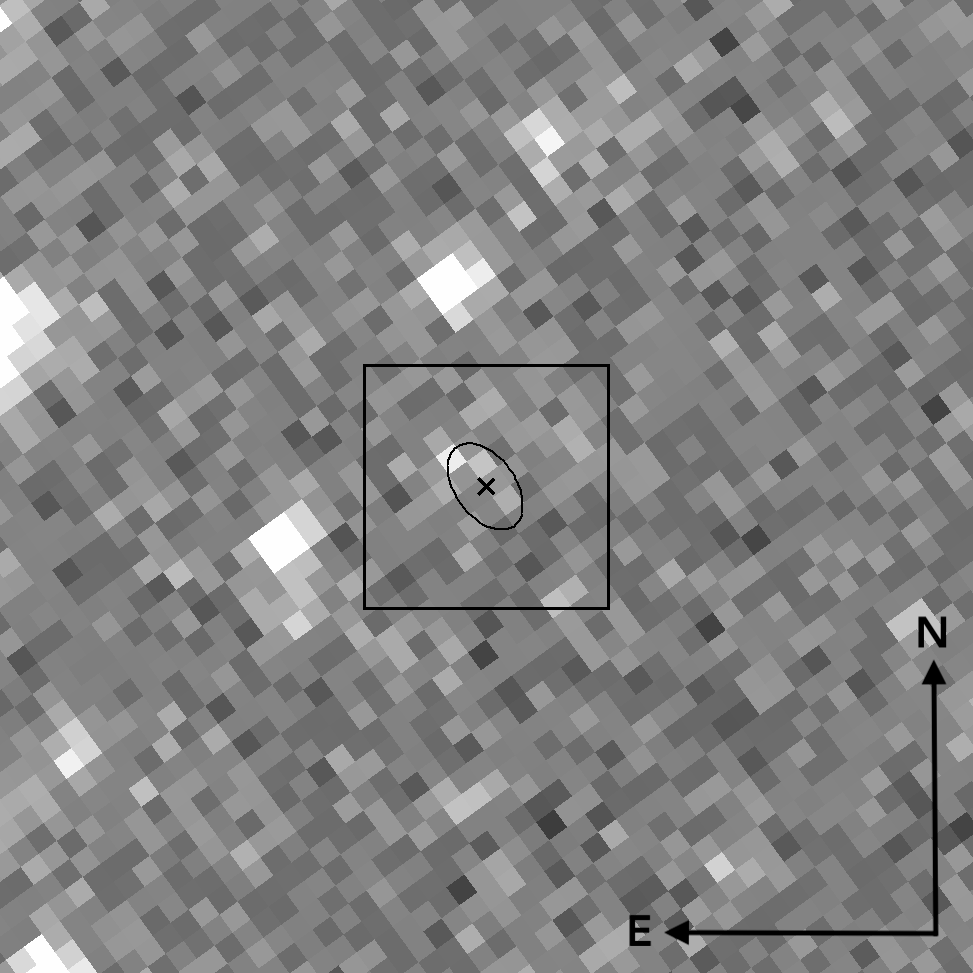}
\label{Progenitor1}
\centering
\includegraphics[width=0.323333\textwidth]{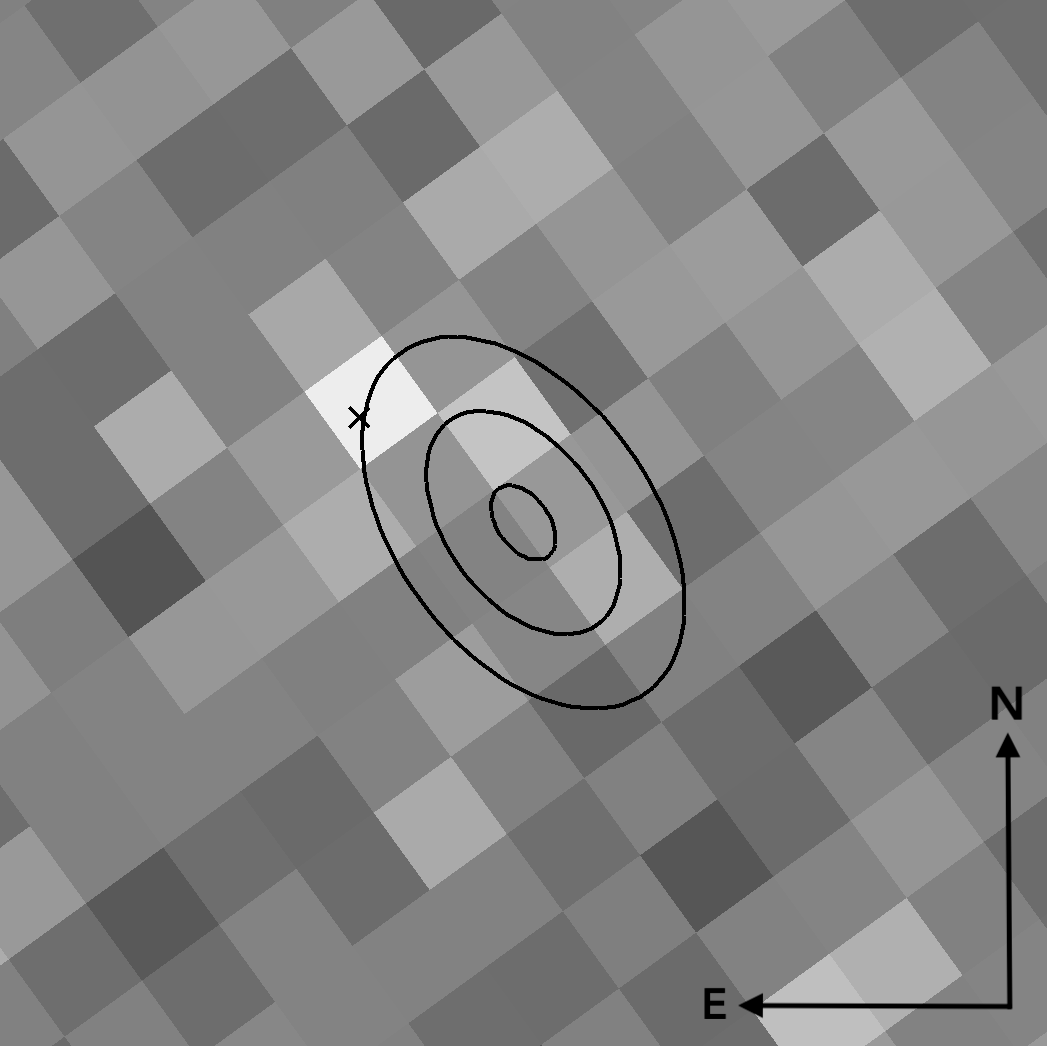}
\label{Progenitor814}
\caption{The location of AT\,2017fvz. Left:\ A $4^{\prime} \times 4^{\prime}$ subset of an LT $V$-band image of AT\,2017fvz taken 49\,d post-eruption. The position of AT\,2017fvz is indicated by the two black lines. Middle:\ A $2^{\prime \prime} \times 2^{\prime \prime}$ subset of the {\it HST} WFPC2 F814W image from April 2007 of the nova field. The black ellipse is the $5\sigma$ uncertainty in the position of the nova (black $\times$). The black box indicates the zoomed-in region in the right image. Right:\ $0^{\prime \prime}\!\!.5 \times 0^{\prime \prime}\!\!.5$ region around AT\,2017fvz in the {\it HST} image. The black ellipses show the 1$\sigma$, 3$\sigma$, and 5$\sigma$ uncertainties on the position of the nova. The black cross is a nearby source not resolvable from the ground.}
\label{Progenitor}
\end{figure*}

As described by \citet{2009ApJ...705.1056B}, \citet{2014A&A...563L...9D}, and in detail by \citet{2014ApJS..213...10W}, we used reference stars in the LT images and an F814W \textit{HST} image to compute a precise astrometric transformation between the datasets. We extended the technique by employing all 18 of the $i'$-band and $r'$-band LT images of AT\,2017fvz to calculate the average nova position (and subsequent scatter) to more precisely and accurately constrain the nova position in the \textit{HST} data, as shown in Figure~\ref{Progenitor}.

We performed crowded-field PSF fitting photometry with DOLPHOT \citep[v2.0;][using standard WFPC2 parameters]{2000PASP..112.1383D} on all detected objects in the \textit{HST} image, recovering a source that is within $5.14\sigma$ (2.05 WFPC2/PC pixels) of AT\,2017fvz, an angular separation of $0^{\prime\prime}\!\!.0931$, or a projected distance of 0.21\,pc (see Figure~\ref{Progenitor} for the position and Table~\ref{HSTPhotometry} for the photometry of the source). While seemingly close, we have no knowledge of the line-of-sight separation of the two objects. A colour-magnitude diagram based on these {\it HST} data was used to determine a limiting magnitude of $m_{\mathrm{F814W}} \approx 23.5$. Using the method described by \citet{2016ApJ...817..143W}, the probability of a coincidental alignment between AT\,2017fvz and this source is 18\%. This does not meet the criterion ($\leq5\%$) employed by \cite{2016ApJ...817..143W} to confirm a likely nova candidate. The astrometric separation indicates with high confidence that this is indeed a chance alignment. The absence of a detected progenitor within the \textit{HST} data indicates that the system is highly likely to harbour a main sequence or subgiant donor, and that the mass accretion rate is modest at best.

\begin{table}
\caption{{\it HST}/WFPC2 photometry of the nearby source.$^\mathrm{a}$}
\label{HSTPhotometry}
\begin{center}
\begin{tabular}{lc}
\hline
\hline
Filter & Photometry (mag) \\
\hline
\hline
F170W & $18.481 \pm 0.480$ \\
F255W & \nodata \\
F336W & $20.982 \pm 0.392$ \\
F439W & \nodata \\
F555W & $23.372 \pm 0.201$ \\
F814W & $22.259 \pm 0.137$ \\
\hline
\hline
\end{tabular}
\end{center}
\begin{tablenotes}
\small
\item $^\mathrm{a}$No source was detected in the F255W or F439W data.
\end{tablenotes}
\end{table}

\begin{figure}
\centering
\includegraphics[width=0.99\columnwidth]{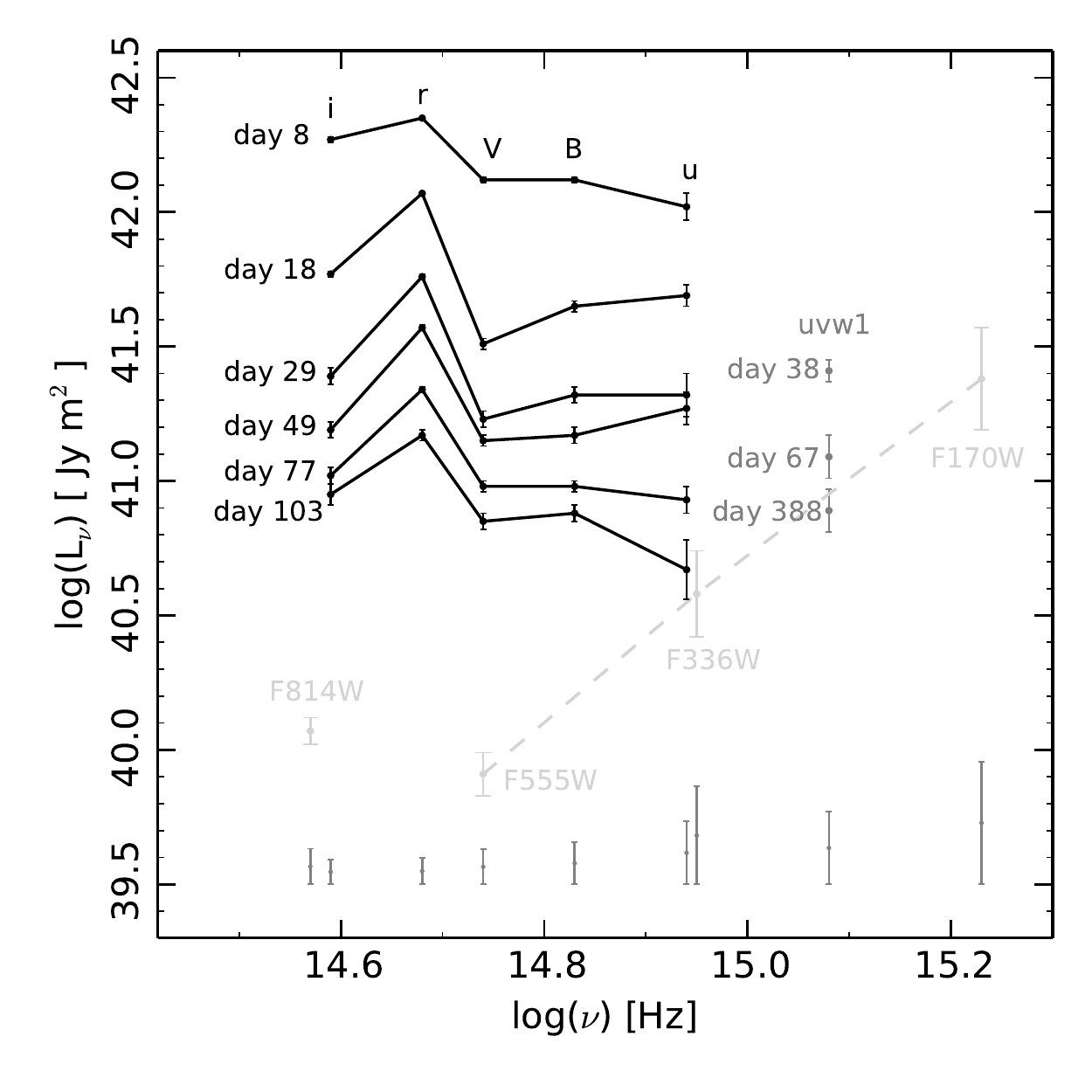}
\caption{SED of AT\,2017fvz (from 8\,d to 103\,d post-eruption) and the source within $0^{\prime\prime}\!\!.1$. Black points are optical photometry of AT\,2017fvz, grey points are from the $uvw1$ photometry. The light grey points are photometry associated with the nearby source, with one of the points being the apparently discrepant F814W photometry. The grey bars at the base are the combined systematic uncertainties from the distance and extinction toward NGC\,6822.}
\label{SED}
\end{figure}

The proximity of this bright source to the nova ($\sim0^{\prime\prime}\!\!.1$) may have contaminated the ground-based and {\it Swift} photometry. Therefore, we determined this source's luminosity in the F814W, F555W, F336W, and F170W filters. Its spectral energy distribution (SED) is shown in Figure~\ref{SED} and compared to the SED evolution of AT\,2017fvz. The source is extremely bright in the NUV, indicating that it is most likely to be an O or B star. The AT\,2017fvz SEDs clearly illustrate the influence that the H$\alpha$ emission of the nova has on the $r'$-band photometry. The final AT\,2017fvz $u'$-band observation ($\sim 103$\,d post-eruption) is consistent with the \textit{HST} F336W photometry (similar wavelengths), indicating that the late-time $u'$ photometry is contaminated by this nearby source. The {\it Swift} photometry is similarly adversely affected. 

A fit to the SED ({\it HST} plus {\it Swift} data) of the nearby source is consistent with the Rayleigh-Jeans tail of a blackbody with $T_\mathrm{eff}=40000\pm8000$\,K and $M\approx-10.1$\,mag ($\chi_\mathrm{red}^{2}=1.68$), at the distance of NGC\,6822 and assuming $E_{B-V}=0.22$\,mag. Such a temperature and luminosity are consistent with an O-star. However, the F814W photometry is significantly brighter than would be expected for such a star.

\section{Discussion}\label{Discussion}

\subsection{The previous nova in NGC 6822}\label{PreviousNova}

There has only been one previous observed nova in NGC\,6822, which was discovered independently by \citet{1999IAUC.7208....3K} and \citet{1999IAUC.7209....2W}. That nova, located at $\alpha=19^\mathrm{h}45^\mathrm{m}0^\mathrm{s}\!.31$, $\delta=-14\degr50^\prime10^{\prime\prime}\!\!.3$ (J2000), was discovered by KAIT in unfiltered images taken on 1999 June 23.40 and 23.44 with $m\approx17.3$\,mag, and by the Beijing Astronomical Observatory Supernova Survey on June 23.69 and 24.72 with an unfiltered magnitude of 18. The nova was then imaged on June 24.38 by LOSS with an unfiltered apparent magnitude of $\sim 17.0$ and by the 1\,m telescope at Sutherland Observatory on June 26.08 and 28.07 with $V$-band apparent magnitudes of $19.0\pm0.1$ and $19.6\pm0.1$, respectively \citep{1999IAUC.7211....3B}. 

\begin{figure*}
\includegraphics[width=0.95\textwidth]{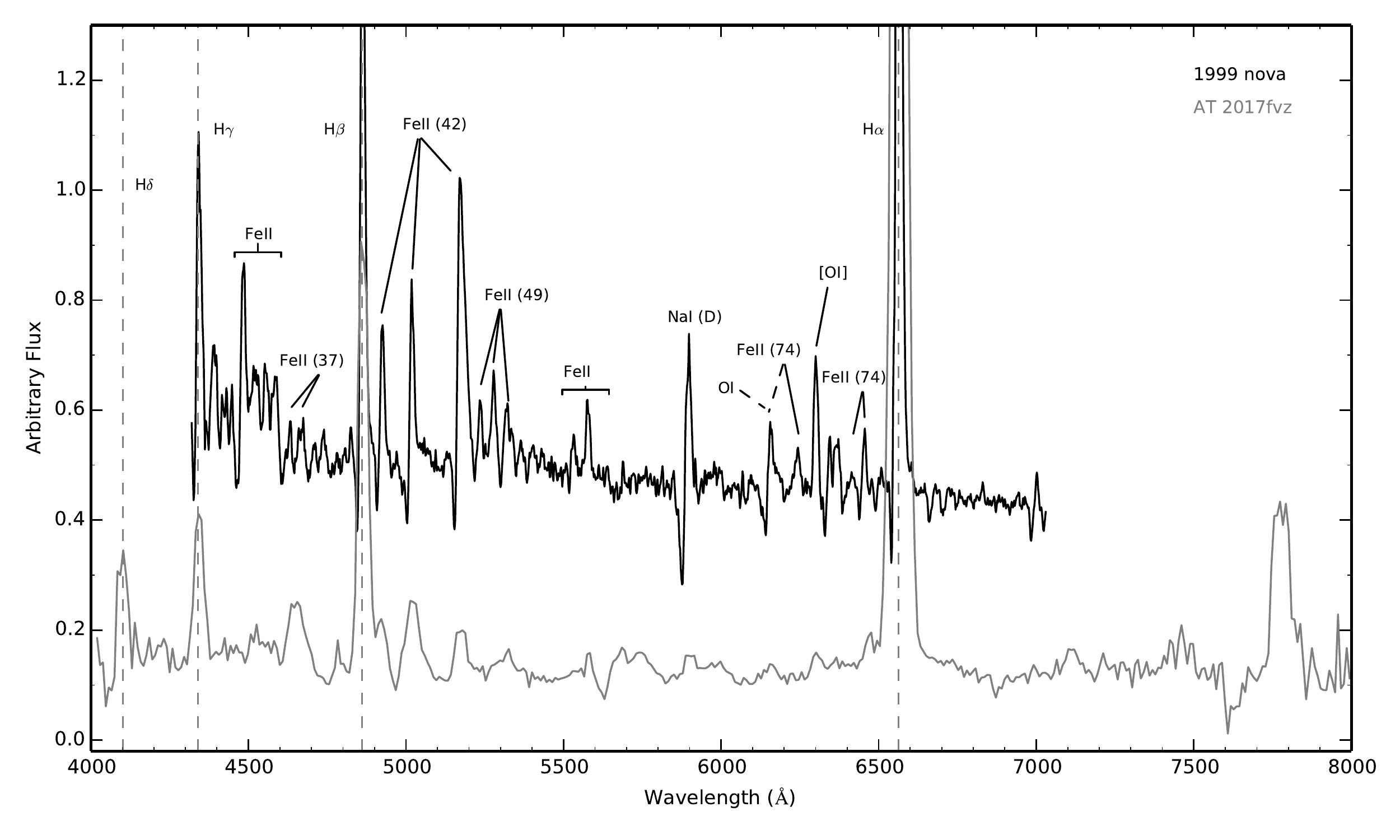}
\caption{Stacked early-time AT\,2017fvz  spectrum ($t = 8$, $14$, $18$\,d) compared to the spectrum of the 1999 nova taken on 1999 July 9 ($\sim16$\,d post-discovery).}
\label{1999Nova}
\end{figure*}

This nova was spectroscopically confirmed on 1999 July 9 using the Kast spectrograph on the 3\,m Shane telescope at Lick Observatory \citep{1999IAUC.7220....2F}. If we assume that the optical peak occurred at discovery, then this spectrum was taken on $t \approx 16$\,d, roughly comparable to the $t \approx 14$\,d and $t \approx 18$\,d spectra of AT\,2017fvz. This spectrum is published for the first time in Figure~\ref{1999Nova} alongside a stacked spectrum of AT\,2017fvz from $t \approx 8$\,d, $t \approx 14$\,d, and $t \approx 18$\,d for comparative purposes.

Just as we see in the spectra of AT\,2017fvz, there are prominent Balmer lines and many of the same \ion{Fe}{ii} lines. Blueward of H$\beta$ there is \ion{Fe}{ii} (37) at 4629\,\AA\ and 4666\,\AA\ and redward there is \ion{Fe}{ii} (42) at 4924\,\AA, 5018\,\AA, and 5169\,\AA. The \ion{Fe}{ii} (74) multiplet is located to the blue of H$\alpha$ at 6148\,\AA, 6248\,\AA, 6417\,\AA, and 6456\,\AA, and again \ion{O}{i} $\lambda$6158 may contribute to the emission line at 6156\,\AA. The \ion{Na}{i}\,D and [\ion{O}{i}] emission lines at 5892\,\AA\ and 6300\,\AA\ (respectively) are also present and much more apparent. As well as the large number of \ion{Fe}{ii} lines between H$\beta$ and H$\gamma$ that were not clearly visible in the AT\,2017fvz spectrum, we see the \ion{Fe}{ii} (49) multiplet at 5235\,\AA, 5276\,\AA, and 5326\,\AA\ to the red of the \ion{Fe}{ii} (42) multiplet. There is a feature in the spectrum at approximately 5533\,\AA\ which may also be \ion{Fe}{ii} and a feature at 5573\,\AA\ which is likely to be [\ion{O}{i}] $\lambda$5577 given the prominence of the [\ion{O}{i}] $\lambda$6300 emission line.

Many of the lines, such as \ion{Na}{i}\,D, \ion{Fe}{ii} (42), and the Balmer lines have P\,Cygni profiles indicating that this spectrum was taken as the nova transitioned from the fireball stage. Comparing directly to the evolution of AT\,2017fvz implies that the 1999 nova evolved more slowly, which is consistent with the much narrower emission lines. See Table~\ref{1999SpecVel} for the emission-line velocities of many of the prominent emission lines with the corresponding velocities for AT\,2017fvz.

\begin{table}
\caption{Comparison of emission-line FWHM velocities (km\,s$^{-1}$).}
\label{1999SpecVel}
\begin{center}
\begin{tabular}{llrr}
\hline
\hline
Line  & Wavelength & 1999 nova & AT\,2017fvz \\
identification & (\AA) & (NGC\,6822) & ($t=14.040$\,d)  \\
\hline
\hline
H$\gamma$ & 4341 & 900 $\pm$ 120 & 2000 $\pm$ 200 \\
\ion{Fe}{ii} (37) & 4491 & 910 $\pm$ 60 & \nodata \\
H$\beta$ & 4861 & 970 $\pm$ 50 & 2100 $\pm$ 210 \\
\ion{Fe}{ii} (42) & 4924 & 840 $\pm$ 50 & \nodata \\
\ion{Fe}{ii} (42) & 5018 & 840 $\pm$ 80 & 2200 $\pm$ 230 \\
\ion{Fe}{ii} (42) & 5169 & 1500 $\pm$ 110 & \nodata \\
\ion{Fe}{ii} (49) & 5235 & 860 $\pm$ 30 & \nodata \\
\ion{Fe}{ii} (49) & 5276 & 1120 $\pm$ 50 & \nodata \\
\ion{Fe}{ii} (49) & 5326 & 1600 $\pm$ 150 & \nodata \\
\ion{Fe}{ii} & 5533 & 840 $\pm$ 90 & \nodata \\
{[\ion{O}{i}]} & 5577 & 850 $\pm$ 60 & \nodata \\
\ion{Na}{i}\,D & 5892 & 800 $\pm$ 130 & \nodata \\
{[\ion{O}{i}]} & 6300 & 650 $\pm$ 40 & \nodata \\
H$\alpha$ & 6563 & 830 $\pm$ 20 & 2300 $\pm$ 100 \\
\hline
\hline
\end{tabular}
\end{center}
\end{table}

\subsection{The lack of X-rays}

We do not detect X-ray emission from AT\,2017fvz in any of the five {\it Swift} observations. This presents two scenarios:\ either the emission was not detectable, or it was detectable but we did not observe the system at the correct time.

There are two reasons why the X-rays emanating from the WD surface may not have been detectable. One option is that the X-ray emission may have ceased before the ejecta surrounding the nova were sufficiently diffuse to permit observation --- that is, $t_{\mathrm{off}} < t_{\mathrm{on}}$.

The alternative is that the SSS may have been too faint to be detected, below the X-ray luminosity upper limit of $\sim10^{37}$\,erg\,s$^{-1}$. A number of SSSs in M\,31 have been particularly faint, but these have been limited to (suspected) slow novae. M31N\,2003-08c had a luminosity of $3.5 \times 10^{36}$\,erg\,s$^{-1}$ when it was first detected $\sim1540$\,d post-eruption and M31N\,2006-09c had a luminosity $\leq 4.0 \times 10^{36}$\,erg\,s$^{-1}$ $\sim 426$\,d post-eruption \citep{2011A&A...533A..52H}. Both lacked photometric data to compute decline times, but we can reasonably assume that they are slow novae owing to their low ejecta velocities. The FWHM of the H$\alpha$ emission line in M31N\,2003-08c is 900\,km\,s$^{-1}$ \citep{2003IAUC.8231....4D} and M31N\,2006-09c has an expansion velocity of $570 \pm 45$\,km\,s$^{-1}$ \citep{2011A&A...533A..52H}. Given their low ejection velocities, the observed turn-on times for these novae are fairly consistent with estimates determined from \citet{2014A&A...563A...2H} for $t_\mathrm{on}$. As such, we would not expect such a late $t_\mathrm{on}$ for AT\,2017fvz. As AT\,2017fvz does not belong to the slow speed class, a faint X-ray luminosity is potentially explained by the low-metallicity environment of NGC\,6822. Depending upon the amount of mixing between the accreted envelope and the underlying WD, the metallicity of the accreted shell will either only weakly (strong mixing) or strongly (little mixing) depend upon the metallicity of the donor.  As the TNR operates via the hot-CNO cycle, a lower metallicity shell might therefore be expected to produce a lower luminosity, but a longer lived SSS phase. In such a scenario, low metallicity alone might explain the lack of any X-ray detection. \citet{2013AstRv...8a..71O} provides further discussion about SSS populations within the SMC, a possibly similar environment to that of NGC\,6822.

Alternatively, if the X-ray emission was in principle detectable, then the reasons for not observing this SSS phase revolve around the timing of the observations. It may also indicate that the \citet{2014A&A...563A...2H} correlations used to predict $t_{\mathrm{on}}$ and $t_{\mathrm{off}}$ (derived from CNe in M\,31) are not valid in the lower metallicity environment of NGC\,6822 \citep[see, e.g.,][]{Williams_2017}. Firstly, the supersoft X-ray source may occur after 388\,d (our last {\it Swift} observation), so we have simply observed too early, indicative of high-mass ejecta and also a low-mass WD. Secondly, the whole SSS phase may have taken place within one of the observing gaps, either between 38\,d and 67\,d, between 67\,d and 97\,d, between 97\,d and 268\,d, or between 243\,d and 388\,d. Though unlikely, there are examples of very short SSS phases in fast novae such as M31N\,2007-12d, which had an extremely short SSS phase of $< 20$\,d \citep{2011A&A...533A..52H}. Finally, the most tantalising option is that the entire SSS phase took place before our first {\it Swift} observation, 38\,d post-eruption. This would imply low-mass ejecta and a high-mass WD, and potentially a recurrent nova.

\subsection{A possibly `faint and fast' or recurrent nova?}\label{PossibleRN}

With $t_{2,V} = 8.1 \pm 0.2$\,d, AT\,2017fvz is a `very-fast' fading nova. We calculated from the MMRD relations of \citet{2000AJ....120.2007D} an expected peak $V$-band absolute magnitude of $M_V \approx -9$, but with a peak absolute magnitude in the range $-7.41>M_V>-8.33$\,mag AT\,2017fvz may be substantially fainter than `expected.' Given this range, and after accounting for expected differences between the $V$-band and $g$ filters \citep[see][]{2009ApJ...690.1148S}, AT\,2017fvz would lie below the MMRD (broadly consistent with the position of M31N\,2008-11a) as presented by \citet[see their Figure~12]{2011ApJ...735...94K}, which plots six `faint and fast' novae by their $t_2$ and their peak absolute $g$-band magnitude. Here we suggest caution, as the upper end of this range (high internal extinction contribution and missed light-curve peak) is marginally consistent with the MMRD. We also note that \citet{2011ApJ...735...94K} employed the Balmer decrement to correct for extinction toward many of their M\,31 novae; however, Case B recombination is not valid in the early stages of nova evolution \citep[see][for a discussion]{Williams_2017}.

The `faint and fast' region of the MMRD phase space is populated by a number of Galactic \citep[see their Figure 13]{2011ApJ...735...94K} and M\,31 RNe. \citet{2014ApJ...788..164P} defined a number of key indicators for a RN masquerading as a CN (i.e., only one observed eruption). AT\,2017fvz satisfies some of these; for example the short $t_2$ implies the presence of a high-mass WD. The high ejecta velocities (for an \ion{Fe}{ii} nova) inferred from the H$\alpha$ emission line ($2430 \pm 70$\,km\,s$^{-1}$) further reinforce this suggestion.

Additionally, there is a plateau in the optical light curve from around day 25 to day 45. It has been proposed that such plateaus are produced by the SSS irradiating a reformed, or surviving, accretion disk and the donor. The subsequent reprocessed optical light then dominates the light emitted by the nova ejecta, temporarily halting the decline of the light curve \citep{2000ApJ...528L..97H,2008ASPC..401.....E,2016ApJ...833..149D}. This could indicate that the accretion disk survived the eruption, pointing to a high accretion rate and/or low ejected mass --- a reasonable indicator of a RN. However, it does not provide strong evidence in isolation. Additionally, the spectrum obtained during the plateau shows no evidence for narrow (or any) \ion{He}{ii} lines, a key signature of a hot disk \citep[as in seen during the plateau phase of known recurrent novae; e.g.,][]{2018ApJ...857...68H}.

The other criteria suggested by \citet{2014ApJ...788..164P} require either far superior spectroscopy or identification of the quiescent system. AT\,2017fvz matches all of their RN indicators that we can reasonably test. The lack of a detected progenitor also indicates the absence of a luminous accretion disk, therefore at most only a modest accretion rate. Even if this system were a RN, it certainly would not be a short-cycle recurrent system.

\section{Summary and Conclusions}\label{Summary and Conclusions}

In this paper we present observations and analysis of AT\,2017fvz, the second nova observed in the Local Group dwarf irregular galaxy NGC\,6822. We carried out detailed photometric and spectroscopic observations of the nova from its initial rise through to the nebular phase. We summarise as follows.

\begin{enumerate}

\item[(1)] AT\,2017fvz was spectroscopically confirmed as an \ion{Fe}{ii} nova, but exhibited broader than typical emission lines.

\item[(2)] The light-curve evolution indicates that AT\,2017fvz may belong to the P-class (plateau) novae, a proposed indication of a surviving or reformed accretion disk.

\item[(3)] As a `very fast' nova with a decline time $t_{2 (V)} \approx 8$\,d, the MMRD predicted peak magnitude is $M_V\approx-9$. Yet we estimate the observed peak is in the range $-7.41>M_V>-8.33$.

\item[(4)] The rapid decline and possible low luminosity suggest that AT\,2017fvz may be a `faint and fast' nova.

\item[(5)] No X-rays were detected between 38 and 388 days post-eruption, therefore the SSS must have occurred within the first $\sim40$\,d, been fully obscured by the ejecta, or simply been too faint to be detectable --- a possible metallicity effect.

\item[(6)] The progenitor system was not recoverable from {\it HST} data, indicating a main sequence or subgiant donor.
\end{enumerate}

We have also included, for the first time, the sparse available data for the other confirmed nova in NGC\,6822. Although currently limited in number, the study of novae across a range of galaxy types will permit systematic studies of how environment --- particularly metallicity --- can affect the properties of novae.

\section*{Acknowledgements}
The authors would like to thank Mike Shara for his role in refereeing the manuscript, and all those that have contributed to the discussion about this object. M.W.H.\ acknowledges a PhD studentship from the UK Science and Technology Facilities Council (STFC). M.J.D.\ received funding from STFC. K.L.P.\ received funding from the UK Space Agency. The work of A.V.F.'s group at UC Berkeley has been generously supported by the TABASGO Foundation, the Christopher R.\ Redlich Fund, and the Miller Institute for Basic Research in Science (UC Berkeley); additional funding was provided by NASA/{\it HST} grant AR-14295 from the Space Telescope Science Institute (STScI), which is operated by the Association of Universities for Research in Astronomy (AURA), Inc., under NASA contract NAS5-26555. 
We thank the staff of the various observatories at which data were obtained. This work made extensive use of the Liverpool Telescope, which is operated by LJMU on the island of La Palma in the Spanish Observatorio del Roque de los Muchachos of the Instituto de Astrofisica de Canarias with financial support from STFC. This work has made use of data from the Asteroid Terrestrial-impact Last Alert System (ATLAS) project. ATLAS is primarily funded to search for new near-Earth asteroids, through NASA grant NN12AR55G issued under the guidance of Lindley Johnson and Kelly Fast. A byproduct of this search is a collection of images and catalogs of the survey area. The ATLAS science products have been made possible through the contributions of the Institute for Astronomy, the University of Hawaii, the Queen's University Belfast, STScI, and Harvard University. KAIT and its ongoing operation were made possible by donations from Sun Microsystems, Inc., the Hewlett-Packard Company, AutoScope Corporation, Lick Observatory, the NSF, the University of California, the Sylvia \& Jim Katzman Foundation, and the TABASGO Foundation. Research at Lick Observatory is partially supported by a generous gift from Google. This research has made use of the APASS database, located at the AAVSO web site. Funding for APASS has been provided by the Robert Martin Ayers Sciences Fund. PyRAF is a product of STScI, which is operated by AURA for NASA.

\bibliographystyle{mnras}
\bibliography{ResearchPapers}

\begin{thebibliography}{}
\makeatletter
\relax
\def\mn@urlcharsother{\let\do\@makeother \do\$\do\&\do\#\do\^\do\_\do\%\do\~}
\def\mn@doi{\begingroup\mn@urlcharsother \@ifnextchar [ {\mn@doi@}
  {\mn@doi@[]}}
\def\mn@doi@[#1]#2{\def\@tempa{#1}\ifx\@tempa\@empty \href
  {http://dx.doi.org/#2} {doi:#2}\else \href {http://dx.doi.org/#2} {#1}\fi
  \endgroup}
\def\mn@eprint#1#2{\mn@eprint@#1:#2::\@nil}
\def\mn@eprint@arXiv#1{\href {http://arxiv.org/abs/#1} {{\tt arXiv:#1}}}
\def\mn@eprint@dblp#1{\href {http://dblp.uni-trier.de/rec/bibtex/#1.xml}
  {dblp:#1}}
\def\mn@eprint@#1:#2:#3:#4\@nil{\def\@tempa {#1}\def\@tempb {#2}\def\@tempc
  {#3}\ifx \@tempc \@empty \let \@tempc \@tempb \let \@tempb \@tempa \fi \ifx
  \@tempb \@empty \def\@tempb {arXiv}\fi \@ifundefined
  {mn@eprint@\@tempb}{\@tempb:\@tempc}{\expandafter \expandafter \csname
  mn@eprint@\@tempb\endcsname \expandafter{\@tempc}}}

\bibitem[\protect\citeauthoryear{{Arp}}{{Arp}}{1956}]{1956AJ.....61...15A}
{Arp} H.~C.,  1956, \mn@doi [\aj] {10.1086/107284}, \href
  {http://adsabs.harvard.edu/abs/1956AJ.....61...15A} {61, 15}

\bibitem[\protect\citeauthoryear{{Aydi} et~al.,}{{Aydi}
  et~al.}{2018}]{2018MNRAS.474.2679A}
{Aydi} E.,  et~al., 2018, \mn@doi [\mnras] {10.1093/mnras/stx2678}, \href
  {http://adsabs.harvard.edu/abs/2018MNRAS.474.2679A} {474, 2679}

\bibitem[\protect\citeauthoryear{{Bakos} \& {PLANET Collaboration}}{{Bakos} \&
  {PLANET Collaboration}}{1999}]{1999IAUC.7211....3B}
{Bakos} G.,  {PLANET Collaboration} 1999, \iaucirc, \href
  {http://adsabs.harvard.edu/abs/1999IAUC.7211....3B} {7211}

\bibitem[\protect\citeauthoryear{{Blackburn}}{{Blackburn}}{1995}]{1995ASPC...77..367B}
{Blackburn} J.~K.,  1995, in {Shaw} R.~A.,  {Payne} H.~E.,   {Hayes} J.~J.~E.,
  eds,  Astronomical Society of the Pacific Conference Series Vol. 77,
  Astronomical Data Analysis Software and Systems IV. p.~367

\bibitem[\protect\citeauthoryear{{Bode}}{{Bode}}{2010}]{2010AN....331..160B}
{Bode} M.~F.,  2010, \mn@doi [Astronomische Nachrichten]
  {10.1002/asna.200911319}, \href
  {http://adsabs.harvard.edu/abs/2010AN....331..160B} {331, 160}

\bibitem[\protect\citeauthoryear{{Bode}, {Darnley}, {Shafter}, {Page},
  {Smirnova}, {Anupama}  \& {Hilton}}{{Bode}
  et~al.}{2009}]{2009ApJ...705.1056B}
{Bode} M.~F.,  {Darnley} M.~J.,  {Shafter} A.~W.,  {Page} K.~L.,  {Smirnova}
  O.,  {Anupama} G.~C.,   {Hilton} T.,  2009, \mn@doi [\apj]
  {10.1088/0004-637X/705/1/1056}, \href
  {http://adsabs.harvard.edu/abs/2009ApJ...705.1056B} {705, 1056}

\bibitem[\protect\citeauthoryear{{Bowen}}{{Bowen}}{1934}]{1934PASP...46..146B}
{Bowen} I.~S.,  1934, \mn@doi [\pasp] {10.1086/124435}, \href
  {http://adsabs.harvard.edu/abs/1934PASP...46..146B} {46, 146}

\bibitem[\protect\citeauthoryear{{Bowen}}{{Bowen}}{1935}]{1935ApJ....81....1B}
{Bowen} I.~S.,  1935, \mn@doi [\apj] {10.1086/143613}, \href
  {http://adsabs.harvard.edu/abs/1935ApJ....81....1B} {81, 1}

\bibitem[\protect\citeauthoryear{{Burrows} et~al.,}{{Burrows}
  et~al.}{2005}]{2005SSRv..120..165B}
{Burrows} D.~N.,  et~al., 2005, \mn@doi [\ssr] {10.1007/s11214-005-5097-2},
  \href {http://adsabs.harvard.edu/abs/2005SSRv..120..165B} {120, 165}

\bibitem[\protect\citeauthoryear{{Czekala} et~al.,}{{Czekala}
  et~al.}{2013}]{2013ApJ...765...57C}
{Czekala} I.,  et~al., 2013, \mn@doi [\apj] {10.1088/0004-637X/765/1/57}, \href
  {http://adsabs.harvard.edu/abs/2013ApJ...765...57C} {765, 57}

\bibitem[\protect\citeauthoryear{{Darnley}, {Ribeiro}, {Bode}, {Hounsell}  \&
  {Williams}}{{Darnley} et~al.}{2012}]{2012ApJ...746...61D}
{Darnley} M.~J.,  {Ribeiro} V.~A.~R.~M.,  {Bode} M.~F.,  {Hounsell} R.~A.,
  {Williams} R.~P.,  2012, \mn@doi [\apj] {10.1088/0004-637X/746/1/61}, \href
  {http://adsabs.harvard.edu/abs/2012ApJ...746...61D} {746, 61}

\bibitem[\protect\citeauthoryear{{Darnley}, {Williams}, {Bode}, {Henze},
  {Ness}, {Shafter}, {Hornoch}  \& {Votruba}}{{Darnley}
  et~al.}{2014}]{2014A&A...563L...9D}
{Darnley} M.~J.,  {Williams} S.~C.,  {Bode} M.~F.,  {Henze} M.,  {Ness} J.-U.,
  {Shafter} A.~W.,  {Hornoch} K.,   {Votruba} V.,  2014, \mn@doi [\aap]
  {10.1051/0004-6361/201423411}, \href
  {http://adsabs.harvard.edu/abs/2014A%26A...563L...9D} {563, L9}

\bibitem[\protect\citeauthoryear{{Darnley} et~al.,}{{Darnley}
  et~al.}{2016}]{2016ApJ...833..149D}
{Darnley} M.~J.,  et~al., 2016, \mn@doi [\apj] {10.3847/1538-4357/833/2/149},
  \href {http://adsabs.harvard.edu/abs/2016ApJ...833..149D} {833, 149}

\bibitem[\protect\citeauthoryear{{Dolphin}}{{Dolphin}}{2000}]{2000PASP..112.1383D}
{Dolphin} A.~E.,  2000, \mn@doi [\pasp] {10.1086/316630}, \href
  {http://adsabs.harvard.edu/abs/2000PASP..112.1383D} {112, 1383}

\bibitem[\protect\citeauthoryear{{Downes} \& {Duerbeck}}{{Downes} \&
  {Duerbeck}}{2000}]{2000AJ....120.2007D}
{Downes} R.~A.,  {Duerbeck} H.~W.,  2000, \mn@doi [\aj] {10.1086/301551}, \href
  {http://adsabs.harvard.edu/abs/2000AJ....120.2007D} {120, 2007}

\bibitem[\protect\citeauthoryear{{Evans}, {Bode}, {O'Brien}  \&
  {Darnley}}{{Evans} et~al.}{2008}]{2008ASPC..401.....E}
{Evans} A.,  {Bode} M.~F.,  {O'Brien} T.~J.,   {Darnley} M.~J.,  eds, 2008, {RS
  Ophiuchi (2006) and the Recurrent Nova Phenomenon}  Astronomical Society of
  the Pacific Conference Series Vol. 401

\bibitem[\protect\citeauthoryear{{Evans} et~al.,}{{Evans}
  et~al.}{2009}]{2009MNRAS.397.1177E}
{Evans} P.~A.,  et~al., 2009, \mn@doi [\mnras]
  {10.1111/j.1365-2966.2009.14913.x}, \href
  {http://adsabs.harvard.edu/abs/2009MNRAS.397.1177E} {397, 1177}

\bibitem[\protect\citeauthoryear{{Filippenko}}{{Filippenko}}{1999}]{1999IAUC.7220....2F}
{Filippenko} A.~V.,  1999, \iaucirc, \href
  {http://adsabs.harvard.edu/abs/1999IAUC.7220....2F} {7220}

\bibitem[\protect\citeauthoryear{{Filippenko}, {Li}, {Treffers}  \&
  {Modjaz}}{{Filippenko} et~al.}{2001}]{2001ASPC..246..121F}
{Filippenko} A.~V.,  {Li} W.~D.,  {Treffers} R.~R.,   {Modjaz} M.,  2001, in
  {Paczynski} B.,  {Chen} W.-P.,   {Lemme} C.,  eds,  Astronomical Society of
  the Pacific Conference Series Vol. 246, IAU Colloq. 183: Small Telescope
  Astronomy on Global Scales. p.~121

\bibitem[\protect\citeauthoryear{{Gaia Collaboration} et~al.,}{{Gaia
  Collaboration} et~al.}{2018}]{2018A&A...616A...1G}
{Gaia Collaboration} et~al., 2018, \mn@doi [\aap]
  {10.1051/0004-6361/201833051}, \href
  {http://adsabs.harvard.edu/abs/2018A%26A...616A...1G} {616, A1}

\bibitem[\protect\citeauthoryear{{Gallart}, {Aparicio}  \& {Vilchez}}{{Gallart}
  et~al.}{1996}]{1996AJ....112.1928G}
{Gallart} C.,  {Aparicio} A.,   {Vilchez} J.~M.,  1996, \mn@doi [\aj]
  {10.1086/118153}, \href {http://adsabs.harvard.edu/abs/1996AJ....112.1928G}
  {112, 1928}

\bibitem[\protect\citeauthoryear{{Ganeshalingam} et~al.,}{{Ganeshalingam}
  et~al.}{2010}]{2010ApJS..190..418G}
{Ganeshalingam} M.,  et~al., 2010, \mn@doi [\apjs]
  {10.1088/0067-0049/190/2/418}, \href
  {http://adsabs.harvard.edu/abs/2010ApJS..190..418G} {190, 418}

\bibitem[\protect\citeauthoryear{{Gehrels} et~al.,}{{Gehrels}
  et~al.}{2004}]{2004ApJ...611.1005G}
{Gehrels} N.,  et~al., 2004, \mn@doi [\apj] {10.1086/422091}, \href
  {http://adsabs.harvard.edu/abs/2004ApJ...611.1005G} {611, 1005}

\bibitem[\protect\citeauthoryear{{Gieren}, {Pietrzy{\'n}ski}, {Nalewajko},
  {Soszy{\'n}ski}, {Bresolin}, {Kudritzki}, {Minniti}  \&
  {Romanowsky}}{{Gieren} et~al.}{2006}]{2006ApJ...647.1056G}
{Gieren} W.,  {Pietrzy{\'n}ski} G.,  {Nalewajko} K.,  {Soszy{\'n}ski} I.,
  {Bresolin} F.,  {Kudritzki} R.-P.,  {Minniti} D.,   {Romanowsky} A.,  2006,
  \mn@doi [\apj] {10.1086/505574}, \href
  {http://adsabs.harvard.edu/abs/2006ApJ...647.1056G} {647, 1056}

\bibitem[\protect\citeauthoryear{{Green} et~al.,}{{Green}
  et~al.}{2018}]{2018MNRAS.478..651G}
{Green} G.~M.,  et~al., 2018, \mn@doi [\mnras] {10.1093/mnras/sty1008}, \href
  {http://adsabs.harvard.edu/abs/2018MNRAS.478..651G} {478, 651}

\bibitem[\protect\citeauthoryear{{Hachisu} \& {Kato}}{{Hachisu} \&
  {Kato}}{2006}]{2006ApJS..167...59H}
{Hachisu} I.,  {Kato} M.,  2006, \mn@doi [\apjs] {10.1086/508063}, \href
  {http://adsabs.harvard.edu/abs/2006ApJS..167...59H} {167, 59}

\bibitem[\protect\citeauthoryear{{Hachisu}, {Kato}, {Kato}  \&
  {Matsumoto}}{{Hachisu} et~al.}{2000}]{2000ApJ...528L..97H}
{Hachisu} I.,  {Kato} M.,  {Kato} T.,   {Matsumoto} K.,  2000, \mn@doi [\apjl]
  {10.1086/312428}, \href {http://adsabs.harvard.edu/abs/2000ApJ...528L..97H}
  {528, L97}

\bibitem[\protect\citeauthoryear{{Harvey}, {Redman}, {Darnley}, {Williams},
  {Berdyugin}, {Piirola}, {Fitzgerald}  \& {O'Connor}}{{Harvey}
  et~al.}{2018}]{2018A&A...611A...3H}
{Harvey} E.~J.,  {Redman} M.~P.,  {Darnley} M.~J.,  {Williams} S.~C.,
  {Berdyugin} A.,  {Piirola} V.~E.,  {Fitzgerald} K.~P.,   {O'Connor} E.~G.~P.,
   2018, \mn@doi [\aap] {10.1051/0004-6361/201731741}, \href
  {http://adsabs.harvard.edu/abs/2018A%26A...611A...3H} {611, A3}

\bibitem[\protect\citeauthoryear{{Heinze} et~al.,}{{Heinze}
  et~al.}{2018}]{2018AJ....156..241H}
{Heinze} A.~N.,  et~al., 2018, \mn@doi [\aj] {10.3847/1538-3881/aae47f}, \href
  {https://ui.adsabs.harvard.edu/\#abs/2018AJ....156..241H} {156, 241}

\bibitem[\protect\citeauthoryear{{Henden}, {Welch}, {Terrell}  \&
  {Levine}}{{Henden} et~al.}{2009}]{2009AAS...21440702H}
{Henden} A.~A.,  {Welch} D.~L.,  {Terrell} D.,   {Levine} S.~E.,  2009, in
  American Astronomical Society Meeting Abstracts \#214. p.~669

\bibitem[\protect\citeauthoryear{{Henze} et~al.,}{{Henze}
  et~al.}{2011}]{2011A&A...533A..52H}
{Henze} M.,  et~al., 2011, \mn@doi [\aap] {10.1051/0004-6361/201015887}, \href
  {http://adsabs.harvard.edu/abs/2011A%26A...533A..52H} {533, A52}

\bibitem[\protect\citeauthoryear{{Henze} et~al.,}{{Henze}
  et~al.}{2014}]{2014A&A...563A...2H}
{Henze} M.,  et~al., 2014, \mn@doi [\aap] {10.1051/0004-6361/201322426}, \href
  {http://adsabs.harvard.edu/abs/2014A%26A...563A...2H} {563, A2}

\bibitem[\protect\citeauthoryear{{Henze} et~al.,}{{Henze}
  et~al.}{2018}]{2018ApJ...857...68H}
{Henze} M.,  et~al., 2018, \mn@doi [\apj] {10.3847/1538-4357/aab6a6}, \href
  {http://adsabs.harvard.edu/abs/2018ApJ...857...68H} {857, 68}

\bibitem[\protect\citeauthoryear{{Hestenes}, {Zheng}  \&
  {Filippenko}}{{Hestenes} et~al.}{2017}]{2017TNSTR.831....1H}
{Hestenes} J.~C.,  {Zheng} W.,   {Filippenko} A.~V.,  2017, Transient Name
  Server Discovery Report, \href
  {http://adsabs.harvard.edu/abs/2017TNSTR.831....1H} {831}

\bibitem[\protect\citeauthoryear{{Iijima} \& {Esenoglu}}{{Iijima} \&
  {Esenoglu}}{2003}]{2003A&A...404..997I}
{Iijima} T.,  {Esenoglu} H.~H.,  2003, \mn@doi [\aap]
  {10.1051/0004-6361:20030528}, \href
  {http://adsabs.harvard.edu/abs/2003A%26A...404..997I} {404, 997}

\bibitem[\protect\citeauthoryear{{Jester} et~al.,}{{Jester}
  et~al.}{2005}]{2005AJ....130..873J}
{Jester} S.,  et~al., 2005, \mn@doi [\aj] {10.1086/432466}, \href
  {http://adsabs.harvard.edu/abs/2005AJ....130..873J} {130, 873}

\bibitem[\protect\citeauthoryear{{Kalberla}, {Burton}, {Hartmann}, {Arnal},
  {Bajaja}, {Morras}  \& {P{\"o}ppel}}{{Kalberla}
  et~al.}{2005}]{2005A&A...440..775K}
{Kalberla} P.~M.~W.,  {Burton} W.~B.,  {Hartmann} D.,  {Arnal} E.~M.,  {Bajaja}
  E.,  {Morras} R.,   {P{\"o}ppel} W.~G.~L.,  2005, \mn@doi [\aap]
  {10.1051/0004-6361:20041864}, \href
  {http://adsabs.harvard.edu/abs/2005A%26A...440..775K} {440, 775}

\bibitem[\protect\citeauthoryear{{Kasliwal}, {Cenko}, {Kulkarni}, {Ofek},
  {Quimby}  \& {Rau}}{{Kasliwal} et~al.}{2011}]{2011ApJ...735...94K}
{Kasliwal} M.~M.,  {Cenko} S.~B.,  {Kulkarni} S.~R.,  {Ofek} E.~O.,  {Quimby}
  R.,   {Rau} A.,  2011, \mn@doi [\apj] {10.1088/0004-637X/735/2/94}, \href
  {http://adsabs.harvard.edu/abs/2011ApJ...735...94K} {735, 94}

\bibitem[\protect\citeauthoryear{{Kayser}}{{Kayser}}{1967}]{1967AJ.....72..134K}
{Kayser} S.~E.,  1967, \mn@doi [\aj] {10.1086/110210}, \href
  {http://adsabs.harvard.edu/abs/1967AJ.....72..134K} {72, 134}

\bibitem[\protect\citeauthoryear{{King} \& {Li}}{{King} \&
  {Li}}{1999}]{1999IAUC.7208....3K}
{King} J.~Y.,  {Li} W.~D.,  1999, \iaucirc, \href
  {http://adsabs.harvard.edu/abs/1999IAUC.7208....3K} {7208}

\bibitem[\protect\citeauthoryear{{Koribalski} et~al.,}{{Koribalski}
  et~al.}{2004}]{2004AJ....128...16K}
{Koribalski} B.~S.,  et~al., 2004, \mn@doi [\aj] {10.1086/421744}, \href
  {http://adsabs.harvard.edu/abs/2004AJ....128...16K} {128, 16}

\bibitem[\protect\citeauthoryear{{Landsman}}{{Landsman}}{1993}]{1993ASPC...52..246L}
{Landsman} W.~B.,  1993, in {Hanisch} R.~J.,  {Brissenden} R.~J.~V.,   {Barnes}
  J.,  eds,  Astronomical Society of the Pacific Conference Series Vol. 52,
  Astronomical Data Analysis Software and Systems II. p.~246

\bibitem[\protect\citeauthoryear{{Larsen}, {Brodie}, {Wasserman}  \&
  {Strader}}{{Larsen} et~al.}{2018}]{2018A&A...613A..56L}
{Larsen} S.~S.,  {Brodie} J.~P.,  {Wasserman} A.,   {Strader} J.,  2018,
  \mn@doi [\aap] {10.1051/0004-6361/201731909}, \href
  {http://adsabs.harvard.edu/abs/2018A%26A...613A..56L} {613, A56}

\bibitem[\protect\citeauthoryear{{Mason}, {Shore}, {De Gennaro Aquino}, {Izzo},
  {Page}  \& {Schwarz}}{{Mason} et~al.}{2018}]{2018ApJ...853...27M}
{Mason} E.,  {Shore} S.~N.,  {De Gennaro Aquino} I.,  {Izzo} L.,  {Page} K.,
  {Schwarz} G.~J.,  2018, \mn@doi [\apj] {10.3847/1538-4357/aaa247}, \href
  {http://adsabs.harvard.edu/abs/2018ApJ...853...27M} {853, 27}

\bibitem[\protect\citeauthoryear{{Massey}, {Armandroff}, {Pyke}, {Patel}  \&
  {Wilson}}{{Massey} et~al.}{1995}]{1995AJ....110.2715M}
{Massey} P.,  {Armandroff} T.~E.,  {Pyke} R.,  {Patel} K.,   {Wilson} C.~D.,
  1995, \mn@doi [\aj] {10.1086/117725}, \href
  {http://adsabs.harvard.edu/abs/1995AJ....110.2715M} {110, 2715}

\bibitem[\protect\citeauthoryear{{Massey}, {Olsen}, {Hodge}, {Jacoby},
  {McNeill}, {Smith}  \& {Strong}}{{Massey} et~al.}{2007}]{2007AJ....133.2393M}
{Massey} P.,  {Olsen} K.~A.~G.,  {Hodge} P.~W.,  {Jacoby} G.~H.,  {McNeill}
  R.~T.,  {Smith} R.~C.,   {Strong} S.~B.,  2007, \mn@doi [\aj]
  {10.1086/513319}, \href {http://adsabs.harvard.edu/abs/2007AJ....133.2393M}
  {133, 2393}

\bibitem[\protect\citeauthoryear{{Mateo}}{{Mateo}}{1998}]{1998ARA&A..36..435M}
{Mateo} M.~L.,  1998, \mn@doi [\araa] {10.1146/annurev.astro.36.1.435}, \href
  {http://adsabs.harvard.edu/abs/1998ARA%26A..36..435M} {36, 435}

\bibitem[\protect\citeauthoryear{{McAlary}, {Madore}, {McGonegal}, {McLaren}
  \& {Welch}}{{McAlary} et~al.}{1983}]{1983ApJ...273..539M}
{McAlary} C.~W.,  {Madore} B.~F.,  {McGonegal} R.,  {McLaren} R.~A.,   {Welch}
  D.~L.,  1983, \mn@doi [\apj] {10.1086/161390}, \href
  {http://adsabs.harvard.edu/abs/1983ApJ...273..539M} {273, 539}

\bibitem[\protect\citeauthoryear{{Mclaughlin}}{{Mclaughlin}}{1945}]{1945PASP...57...69M}
{Mclaughlin} D.~B.,  1945, \mn@doi [\pasp] {10.1086/125689}, \href
  {http://adsabs.harvard.edu/abs/1945PASP...57...69M} {57, 69}

\bibitem[\protect\citeauthoryear{{Moore}}{{Moore}}{1945}]{1945CoPri..21....1M}
{Moore} C.~E.,  1945, Contributions from the Princeton University Observatory,
  \href {http://adsabs.harvard.edu/abs/1945CoPri..21....1M} {21, 1}

\bibitem[\protect\citeauthoryear{{Orio}}{{Orio}}{2013}]{2013AstRv...8a..71O}
{Orio} M.,  2013, \mn@doi [The Astronomical Review]
  {10.1080/21672857.2013.11519714}, \href
  {http://adsabs.harvard.edu/abs/2013AstRv...8a..71O} {8, 71}

\bibitem[\protect\citeauthoryear{{Osborne} et~al.,}{{Osborne}
  et~al.}{2011}]{2011ApJ...727..124O}
{Osborne} J.~P.,  et~al., 2011, \mn@doi [\apj] {10.1088/0004-637X/727/2/124},
  \href {http://adsabs.harvard.edu/abs/2011ApJ...727..124O} {727, 124}

\bibitem[\protect\citeauthoryear{{Osterbrock} \& {Ferland}}{{Osterbrock} \&
  {Ferland}}{2006}]{2006agna.book.....O}
{Osterbrock} D.~E.,  {Ferland} G.~J.,  2006, {Astrophysics of gaseous nebulae
  and active galactic nuclei}.
{CA: University Science Books}

\bibitem[\protect\citeauthoryear{{Pagnotta} \& {Schaefer}}{{Pagnotta} \&
  {Schaefer}}{2014}]{2014ApJ...788..164P}
{Pagnotta} A.,  {Schaefer} B.~E.,  2014, \mn@doi [\apj]
  {10.1088/0004-637X/788/2/164}, \href
  {http://adsabs.harvard.edu/abs/2014ApJ...788..164P} {788, 164}

\bibitem[\protect\citeauthoryear{{Payne-Gaposchkin}}{{Payne-Gaposchkin}}{1957}]{1957gano.book.....G}
{Payne-Gaposchkin} C.,  1957, {The Galactic Novae}.
{Amsterdam, North-Holland Pub.~Co.}

\bibitem[\protect\citeauthoryear{{Piascik}, {Steele}, {Bates}, {Mottram},
  {Smith}, {Barnsley}  \& {Bolton}}{{Piascik}
  et~al.}{2014}]{2014SPIE.9147E..8HP}
{Piascik} A.~S.,  {Steele} I.~A.,  {Bates} S.~D.,  {Mottram} C.~J.,  {Smith}
  R.~J.,  {Barnsley} R.~M.,   {Bolton} B.,  2014, in Ground-based and Airborne
  Instrumentation for Astronomy V. p. 91478H, \mn@doi{10.1117/12.2055117}

\bibitem[\protect\citeauthoryear{{Rich}, {Persson}, {Freedman}, {Madore},
  {Monson}, {Scowcroft}  \& {Seibert}}{{Rich}
  et~al.}{2014}]{2014ApJ...794..107R}
{Rich} J.~A.,  {Persson} S.~E.,  {Freedman} W.~L.,  {Madore} B.~F.,  {Monson}
  A.~J.,  {Scowcroft} V.,   {Seibert} M.,  2014, \mn@doi [\apj]
  {10.1088/0004-637X/794/2/107}, \href
  {http://adsabs.harvard.edu/abs/2014ApJ...794..107R} {794, 107}

\bibitem[\protect\citeauthoryear{{Roming} et~al.,}{{Roming}
  et~al.}{2005}]{2005SSRv..120...95R}
{Roming} P.~W.~A.,  et~al., 2005, \mn@doi [\ssr] {10.1007/s11214-005-5095-4},
  \href {http://adsabs.harvard.edu/abs/2005SSRv..120...95R} {120, 95}

\bibitem[\protect\citeauthoryear{{Schaefer}}{{Schaefer}}{2018}]{2018MNRAS.481.3033S}
{Schaefer} B.~E.,  2018, \mn@doi [\mnras] {10.1093/mnras/sty2388}, \href
  {http://adsabs.harvard.edu/abs/2018MNRAS.481.3033S} {481, 3033}

\bibitem[\protect\citeauthoryear{{Selvelli} \& {Gilmozzi}}{{Selvelli} \&
  {Gilmozzi}}{2019}]{2019A&A...622A.186S}
{Selvelli} P.,  {Gilmozzi} R.,  2019, \mn@doi [\aap]
  {10.1051/0004-6361/201834238}, \href
  {http://adsabs.harvard.edu/abs/2019A%26A...622A.186S} {622, A186}

\bibitem[\protect\citeauthoryear{{Shafter}}{{Shafter}}{2013}]{2013AJ....145..117S}
{Shafter} A.~W.,  2013, \mn@doi [\aj] {10.1088/0004-6256/145/5/117}, \href
  {http://adsabs.harvard.edu/abs/2013AJ....145..117S} {145, 117}

\bibitem[\protect\citeauthoryear{{Shafter}, {Rau}, {Quimby}, {Kasliwal},
  {Bode}, {Darnley}  \& {Misselt}}{{Shafter}
  et~al.}{2009}]{2009ApJ...690.1148S}
{Shafter} A.~W.,  {Rau} A.,  {Quimby} R.~M.,  {Kasliwal} M.~M.,  {Bode} M.~F.,
  {Darnley} M.~J.,   {Misselt} K.~A.,  2009, \mn@doi [\apj]
  {10.1088/0004-637X/690/2/1148}, \href
  {http://adsabs.harvard.edu/abs/2009ApJ...690.1148S} {690, 1148}

\bibitem[\protect\citeauthoryear{{Shafter} et~al.,}{{Shafter}
  et~al.}{2011}]{2011ApJ...734...12S}
{Shafter} A.~W.,  et~al., 2011, \mn@doi [\apj] {10.1088/0004-637X/734/1/12},
  \href {http://adsabs.harvard.edu/abs/2011ApJ...734...12S} {734, 12}

\bibitem[\protect\citeauthoryear{{Shafter}, {Darnley}, {Bode}  \&
  {Ciardullo}}{{Shafter} et~al.}{2012}]{2012ApJ...752..156S}
{Shafter} A.~W.,  {Darnley} M.~J.,  {Bode} M.~F.,   {Ciardullo} R.,  2012,
  \mn@doi [\apj] {10.1088/0004-637X/752/2/156}, \href
  {http://adsabs.harvard.edu/abs/2012ApJ...752..156S} {752, 156}

\bibitem[\protect\citeauthoryear{{Shafter}, {Curtin}, {Pritchet}, {Bode}  \&
  {Darnley}}{{Shafter} et~al.}{2014}]{2014ASPC..490...77S}
{Shafter} A.~W.,  {Curtin} C.,  {Pritchet} C.~J.,  {Bode} M.~F.,   {Darnley}
  M.~J.,  2014, in {Woudt} P.~A.,  {Ribeiro} V.~A.~R.~M.,  eds,  Astronomical
  Society of the Pacific Conference Series Vol. 490, Stellar Novae: Past and
  Future Decades. p.~77 (\mn@eprint {arXiv} {1307.2296})

\bibitem[\protect\citeauthoryear{{Shappee} et~al.,}{{Shappee}
  et~al.}{2014}]{2014ApJ...788...48S}
{Shappee} B.~J.,  et~al., 2014, \mn@doi [\apj] {10.1088/0004-637X/788/1/48},
  \href {http://adsabs.harvard.edu/abs/2014ApJ...788...48S} {788, 48}

\bibitem[\protect\citeauthoryear{{Shara} et~al.,}{{Shara}
  et~al.}{2016}]{2016ApJS..227....1S}
{Shara} M.~M.,  et~al., 2016, \mn@doi [\apjs] {10.3847/0067-0049/227/1/1},
  \href {http://adsabs.harvard.edu/abs/2016ApJS..227....1S} {227, 1}

\bibitem[\protect\citeauthoryear{{Shara} et~al.,}{{Shara}
  et~al.}{2017}]{2017ApJ...839..109S}
{Shara} M.~M.,  et~al., 2017, \mn@doi [\apj] {10.3847/1538-4357/aa65cd}, \href
  {http://adsabs.harvard.edu/abs/2017ApJ...839..109S} {839, 109}

\bibitem[\protect\citeauthoryear{{Stalder} et~al.,}{{Stalder}
  et~al.}{2017}]{2017ApJ...850..149S}
{Stalder} B.,  et~al., 2017, \mn@doi [\apj] {10.3847/1538-4357/aa95c1}, \href
  {http://adsabs.harvard.edu/abs/2017ApJ...850..149S} {850, 149}

\bibitem[\protect\citeauthoryear{{Starrfield}, {Sparks}  \&
  {Truran}}{{Starrfield} et~al.}{1976}]{1976IAUS...73..155S}
{Starrfield} S.,  {Sparks} W.~M.,   {Truran} J.~W.,  1976, in {Eggleton} P.,
  {Mitton} S.,   {Whelan} J.,  eds,  IAU Symposium Vol. 73, Structure and
  Evolution of Close Binary Systems. pp 155--172

\bibitem[\protect\citeauthoryear{{Steele} et~al.,}{{Steele}
  et~al.}{2004}]{2004SPIE.5489..679S}
{Steele} I.~A.,  et~al., 2004, in {Oschmann} Jr. J.~M.,  ed.,  \procspie Vol.
  5489, Ground-based Telescopes. pp 679--692, \mn@doi{10.1117/12.551456}

\bibitem[\protect\citeauthoryear{{Stetson}}{{Stetson}}{1987}]{1987PASP...99..191S}
{Stetson} P.~B.,  1987, \mn@doi [\pasp] {10.1086/131977}, \href
  {http://adsabs.harvard.edu/abs/1987PASP...99..191S} {99, 191}

\bibitem[\protect\citeauthoryear{{Strope}, {Schaefer}  \& {Henden}}{{Strope}
  et~al.}{2010}]{2010AJ....140...34S}
{Strope} R.~J.,  {Schaefer} B.~E.,   {Henden} A.~A.,  2010, \mn@doi [\aj]
  {10.1088/0004-6256/140/1/34}, \href
  {http://adsabs.harvard.edu/abs/2010AJ....140...34S} {140, 34}

\bibitem[\protect\citeauthoryear{{Tonry} et~al.,}{{Tonry}
  et~al.}{2018}]{2018PASP..130f4505T}
{Tonry} J.~L.,  et~al., 2018, \mn@doi [\pasp] {10.1088/1538-3873/aabadf}, \href
  {http://adsabs.harvard.edu/abs/2018PASP..130f4505T} {130, 064505}

\bibitem[\protect\citeauthoryear{{Walker}}{{Walker}}{1954}]{1954PASP...66..230W}
{Walker} M.~F.,  1954, \mn@doi [\pasp] {10.1086/126703}, \href
  {http://adsabs.harvard.edu/abs/1954PASP...66..230W} {66, 230}

\bibitem[\protect\citeauthoryear{{Warner}}{{Warner}}{1995}]{1995CAS....28.....W}
{Warner} B.,  1995, Cambridge Astrophysics Series, \href
  {http://adsabs.harvard.edu/abs/1995CAS....28.....W} {28}

\bibitem[\protect\citeauthoryear{{Wei}, {Xu}, {Qiao}, {Qiu}  \& {Hu}}{{Wei}
  et~al.}{1999}]{1999IAUC.7209....2W}
{Wei} J.~Y.,  {Xu} D.~W.,  {Qiao} Q.~Y.,  {Qiu} Y.~L.,   {Hu} J.~Y.,  1999,
  \iaucirc, \href {http://adsabs.harvard.edu/abs/1999IAUC.7209....2W} {7209}

\bibitem[\protect\citeauthoryear{{Weldrake}, {de Blok}  \& {Walter}}{{Weldrake}
  et~al.}{2003}]{2003MNRAS.340...12W}
{Weldrake} D.~T.~F.,  {de Blok} W.~J.~G.,   {Walter} F.,  2003, \mn@doi
  [\mnras] {10.1046/j.1365-8711.2003.06170.x}, \href
  {http://adsabs.harvard.edu/abs/2003MNRAS.340...12W} {340, 12}

\bibitem[\protect\citeauthoryear{{Williams}}{{Williams}}{1992}]{1992AJ....104..725W}
{Williams} R.~E.,  1992, \mn@doi [\aj] {10.1086/116268}, \href
  {http://adsabs.harvard.edu/abs/1992AJ....104..725W} {104, 725}

\bibitem[\protect\citeauthoryear{{Williams}}{{Williams}}{1994}]{1994ApJ...426..279W}
{Williams} R.~E.,  1994, \mn@doi [\apj] {10.1086/174062}, \href
  {http://adsabs.harvard.edu/abs/1994ApJ...426..279W} {426, 279}

\bibitem[\protect\citeauthoryear{{Williams}}{{Williams}}{2012}]{2012AJ....144...98W}
{Williams} R.,  2012, \mn@doi [\aj] {10.1088/0004-6256/144/4/98}, \href
  {http://adsabs.harvard.edu/abs/2012AJ....144...98W} {144, 98}

\bibitem[\protect\citeauthoryear{{Williams} \& {Darnley}}{{Williams} \&
  {Darnley}}{2017}]{2017ATel10630....1W}
{Williams} S.~C.,  {Darnley} M.~J.,  2017, The Astronomer's Telegram, \href
  {http://adsabs.harvard.edu/abs/2017ATel10630....1W} {10630}

\bibitem[\protect\citeauthoryear{{Williams}, {Darnley}, {Bode}, {Keen}  \&
  {Shafter}}{{Williams} et~al.}{2014}]{2014ApJS..213...10W}
{Williams} S.~C.,  {Darnley} M.~J.,  {Bode} M.~F.,  {Keen} A.,   {Shafter}
  A.~W.,  2014, \mn@doi [\apjs] {10.1088/0067-0049/213/1/10}, \href
  {http://adsabs.harvard.edu/abs/2014ApJS..213...10W} {213, 10}

\bibitem[\protect\citeauthoryear{{Williams}, {Darnley}, {Bode}  \&
  {Shafter}}{{Williams} et~al.}{2016}]{2016ApJ...817..143W}
{Williams} S.~C.,  {Darnley} M.~J.,  {Bode} M.~F.,   {Shafter} A.~W.,  2016,
  \mn@doi [\apj] {10.3847/0004-637X/817/2/143}, \href
  {http://adsabs.harvard.edu/abs/2016ApJ...817..143W} {817, 143}

\bibitem[\protect\citeauthoryear{Williams, Darnley  \& Henze}{Williams
  et~al.}{2017}]{Williams_2017}
Williams S.~C.,  Darnley M.~J.,   Henze M.,  2017, \mn@doi [\mnras]
  {10.1093/mnras/stx1793}, 472, 1300

\bibitem[\protect\citeauthoryear{{Yaron}, {Prialnik}, {Shara}  \&
  {Kovetz}}{{Yaron} et~al.}{2005}]{2005ApJ...623..398Y}
{Yaron} O.,  {Prialnik} D.,  {Shara} M.~M.,   {Kovetz} A.,  2005, \mn@doi
  [\apj] {10.1086/428435}, \href
  {http://adsabs.harvard.edu/abs/2005ApJ...623..398Y} {623, 398}

\bibitem[\protect\citeauthoryear{{Zwicky}}{{Zwicky}}{1936}]{1936PASP...48..191Z}
{Zwicky} F.,  1936, \mn@doi [\pasp] {10.1086/124698}, \href
  {http://adsabs.harvard.edu/abs/1936PASP...48..191Z} {48, 191}

\bibitem[\protect\citeauthoryear{{di Mille}, {Ciroi}, {Botte}  \&
  {Boschetti}}{{di Mille} et~al.}{2003}]{2003IAUC.8231....4D}
{di Mille} F.,  {Ciroi} S.,  {Botte} V.,   {Boschetti} C.~S.,  2003, \iaucirc,
  \href {http://adsabs.harvard.edu/abs/2003IAUC.8231....4D} {8231}

\bibitem[\protect\citeauthoryear{{van den Heuvel}, {Bhattacharya}, {Nomoto}  \&
  {Rappaport}}{{van den Heuvel} et~al.}{1992}]{1992A&A...262...97V}
{van den Heuvel} E.~P.~J.,  {Bhattacharya} D.,  {Nomoto} K.,   {Rappaport}
  S.~A.,  1992, \aap, \href
  {http://adsabs.harvard.edu/abs/1992A%26A...262...97V} {262, 97}

\makeatother
\end{thebibliography}
\balance

\appendix

\section{Photometry of AT 2017fvz}

In Table~\ref{Photometry1} we provide the optical photometry of AT\,2017fvz from the Liverpool Telescope and ASAS-SN. Table~\ref{Photometry2} lists the photometry of AT\,2017fvz from KAIT. Table~\ref{Photometry3} presents the ATLAS photometry of AT\,2017fvz.

\begin{table*}
\caption{LT and ASAS-SN optical photometry of AT\,2017fvz.}
\centering
\label{Photometry1}
\begin{tabular}{l c r c c c r}
\hline
\hline
UT Date & MJD (d) & $t-t_0$ (d) & Telescope \& instrument & Exposure time (s) & Filter & Photometry (mag) \\
\hline
\hline
2017-08-09.916 & 57974.916 & 8.032 & LT IO:O & 60 & $u'$ & 18.522 $\pm$ 0.118 \\
2017-08-15.908 & 57980.908 & 14.024 & LT IO:O & 60 & $u'$ & 18.737 $\pm$ 0.063 \\
2017-08-17.894 & 57982.894 & 16.010 & LT IO:O & 60 & $u'$ & 19.248 $\pm$ 0.131 \\
2017-08-19.910 & 57984.910 & 18.026 & LT IO:O & 60 & $u'$ & 19.337 $\pm$ 0.111 \\
2017-08-23.889 & 57988.889 & 22.005 & LT IO:O & 60 & $u'$ & 19.633 $\pm$ 0.127 \\
2017-08-30.883 & 57995.883 & 28.999 & LT IO:O & 120 & $u'$ & 20.265 $\pm$ 0.191 \\
2017-09-04.939 & 58000.939 & 34.055 & LT IO:O & 120 & $u'$ & 20.722 $\pm$ 0.335 \\
2017-09-19.936 & 58015.936 & 49.052 & LT IO:O & 120 & $u'$ & 20.378 $\pm$ 0.158 \\
2017-09-24.896 & 58020.896 & 54.012 & LT IO:O & 120 & $u'$ & 20.571 $\pm$ 0.274 \\
2017-10-17.842 & 58043.842 & 76.958 & LT IO:O & 120 & $u'$ & 21.228 $\pm$ 0.131 \\
2017-11-12.822 & 58098.822 & 102.938 & LT IO:O & 120 & $u'$ & 21.895 $\pm$ 0.270 \\
\hline
2017-08-09.917 & 57974.917 & 8.033 & LT IO:O & 60 & $B$ & 18.224 $\pm$ 0.019 \\
2017-08-15.910 & 57980.910 & 14.026 & LT IO:O & 60 & $B$ & 19.003 $\pm$ 0.025 \\
2017-08-17.895 & 57982.895 & 16.011 & LT IO:O & 60 & $B$ & 19.198 $\pm$ 0.032 \\
2017-08-19.911 & 57984.911 & 18.027 & LT IO:O & 60 & $B$ & 19.380 $\pm$ 0.038 \\
2017-08-23.890 & 57988.890 & 22.006 & LT IO:O & 60 & $B$ & 19.903 $\pm$ 0.049 \\
2017-08-30.885 & 57995.885 & 29.001 & LT IO:O & 120 & $B$ & 20.204 $\pm$ 0.086 \\
2017-09-04.941 & 58000.941 & 34.057 & LT IO:O & 120 & $B$ & 19.987 $\pm$ 0.081 \\
2017-09-19.941 & 58015.941 & 49.057 & LT IO:O & 120 & $B$ & 20.591 $\pm$ 0.072 \\
2017-09-24.901 & 58020.901 & 54.017 & LT IO:O & 120 & $B$ & 20.875 $\pm$ 0.106 \\
2017-10-17.847 & 58043.847 & 76.963 & LT IO:O & 120 & $B$ & 21.066 $\pm$ 0.059 \\
2017-11-12.827 & 58098.827 & 102.943 & LT IO:O & 120 & $B$ & 21.321 $\pm$ 0.071 \\
\hline
2017-08-03.190 & 57968.190 & 1.306 & ASAS-SN & 270 & $V$ & 16.654 \\
2017-08-09.918 & 57974.918 & 8.034 & LT IO:O & 60 & $V$ & 17.905 $\pm$ 0.017 \\
2017-08-15.911 & 57980.911 & 14.027 & LT IO:O & 60 & $V$ & 18.722 $\pm$ 0.017 \\
2017-08-17.896 & 57982.896 & 16.012 & LT IO:O & 60 & $V$ & 19.035 $\pm$ 0.029 \\
2017-08-19.912 & 57984.912 & 18.028 & LT IO:O & 60 & $V$ & 19.428 $\pm$ 0.045 \\
2017-08-23.891 & 57988.891 & 22.007 & LT IO:O & 60 & $V$ & 19.691 $\pm$ 0.048 \\
2017-08-30.887 & 57995.887 & 29.003 & LT IO:O & 120 & $V$ & 20.128 $\pm$ 0.073 \\
2017-09-04.943 & 58000.943 & 34.059 & LT IO:O & 120 & $V$ & 20.148 $\pm$ 0.097 \\
2017-09-19.946 & 58015.946 & 49.062 & LT IO:O & 120 & $V$ & 20.341 $\pm$ 0.061 \\
2017-09-24.906 & 58020.906 & 54.022 & LT IO:O & 120 & $V$ & 20.565 $\pm$ 0.092 \\
2017-10-17.852 & 58043.852 & 76.968 & LT IO:O & 120 & $V$ & 20.758 $\pm$ 0.056 \\
2017-11-12.832 & 58098.832 & 102.948 & LT IO:O & 120 & $V$ & 21.079 $\pm$ 0.077 \\
\hline
2017-08-09.919 & 57974.919 & 8.035 & LT IO:O & 60 & $r'$ & 17.186 $\pm$ 0.011 \\
2017-08-15.912 & 57980.912 & 14.028 & LT IO:O & 60 & $r'$ & 17.589 $\pm$ 0.009 \\
2017-08-17.897 & 57982.897 & 16.013 & LT IO:O & 60 & $r'$ & 17.729 $\pm$ 0.010 \\
2017-08-19.913 & 57984.913 & 18.029 & LT IO:O & 60 & $r'$ & 17.886 $\pm$ 0.011 \\
2017-08-23.892 & 57988.892 & 22.008 & LT IO:O & 60 & $r'$ & 18.212 $\pm$ 0.013 \\
2017-08-30.889 & 57995.889 & 29.005 & LT IO:O & 120 & $r'$ & 18.681 $\pm$ 0.023 \\
2017-09-04.944 & 58000.944 & 34.060 & LT IO:O & 120 & $r'$ & 18.750 $\pm$ 0.024 \\
2017-09-19.951 & 58015.951 & 49.067 & LT IO:O & 60 & $r'$ & 19.153 $\pm$ 0.028 \\
2017-09-24.911 & 58020.911 & 54.027 & LT IO:O & 60 & $r'$ & 19.255 $\pm$ 0.038 \\
2017-10-17.857 & 58043.857 & 76.973 & LT IO:O & 60 & $r'$ & 19.729 $\pm$ 0.030 \\
2017-11-12.837 & 58098.837 & 102.953 & LT IO:O & 60 & $r'$ & 20.143 $\pm$ 0.046 \\
\hline
2017-08-09.890 & 57974.890 & 8.006 & LT SPRAT & 10 & $r'$ & 17.391 $\pm$ 0.027 \\
2017-08-15.915 & 57980.915 & 14.031 & LT SPRAT & 10 & $r'$ & 17.983 $\pm$ 0.031 \\
2017-08-19.886 & 57984.886 & 18.002 & LT SPRAT & 10 & $r'$ & 18.268 $\pm$ 0.035 \\
2017-08-25.874 & 57990.874 & 23.990 & LT SPRAT & 10 & $r'$ & 17.951 $\pm$ 0.174 \\
2017-09-11.884 & 58007.884 & 41.000 & LT SPRAT & 10 & $r'$ & 19.286 $\pm$ 0.043 \\
2017-09-12.865 & 58008.865 & 41.981 & LT SPRAT & 10 & $r'$ & 19.368 $\pm$ 0.050 \\
2017-10-10.834 & 58036.834 & 69.950 & LT SPRAT & 10 & $r'$ & 20.040 $\pm$ 0.067 \\
\hline
2017-08-09.920 & 57974.920 & 8.036 & LT IO:O & 60 & $i'$ & 17.246 $\pm$ 0.021 \\
2017-08-15.913 & 57980.913 & 14.029 & LT IO:O & 60 & $i'$ & 18.044 $\pm$ 0.017 \\
2017-08-17.898 & 57982.898 & 16.014 & LT IO:O & 60 & $i'$ & 18.230 $\pm$ 0.023 \\
2017-08-19.914 & 57984.914 & 18.030 & LT IO:O & 60 & $i'$ & 18.496 $\pm$ 0.026 \\
2017-08-23.893 & 57988.893 & 22.009 & LT IO:O & 60 & $i'$ & 18.913 $\pm$ 0.030 \\
2017-08-30.890 & 57995.890 & 29.006 & LT IO:O & 120 & $i'$ & 19.449 $\pm$ 0.068 \\
2017-09-04.946 & 58000.946 & 34.062 & LT IO:O & 120 & $i'$ & 19.569 $\pm$ 0.050 \\
2017-09-19.954 & 58015.954 & 49.070 & LT IO:O & 60 & $i'$ & 19.957 $\pm$ 0.065 \\
2017-09-24.914 & 58020.914 & 54.030 & LT IO:O & 60 & $i'$ & 19.649 $\pm$ 0.082 \\
2017-10-17.860 & 58043.860 & 76.976 & LT IO:O & 60 & $i'$ & 20.378 $\pm$ 0.076 \\
2017-11-12.840 & 58098.840 & 102.956 & LT IO:O & 60 & $i'$ & 20.547 $\pm$ 0.095 \\
\hline
\hline
\end{tabular}
\end{table*}

\begin{table*}
\caption{KAIT photometry of AT\,2017fvz. The clear filter is treated as the $r'$-band.}
\centering
\label{Photometry2}
\begin{tabular}{l c r c c r}
\hline
\hline
UT Date & MJD (d) & $t-t_0$ (d) & Telescope \& instrument & Filter & Photometry (mag) \\
\hline
\hline
2017-08-02.384 & 57967.384 & 0.500 & KAIT & Clear & 17.61 $\pm$ 0.09 \\
2017-08-03.289 & 57968.289 & 1.405 & KAIT & Clear & 16.44 $\pm$ 0.11 \\
2017-08-08.211 & 57973.211 & 6.327 & KAIT & Clear & 16.83 $\pm$ 0.04 \\
2017-08-09.293 & 57974.293 & 7.409 & KAIT & Clear & 16.87 $\pm$ 0.06 \\
2017-08-10.370 & 57975.370 & 8.486 & KAIT & Clear & 17.04 $\pm$ 0.06 \\
2017-08-11.366 & 57976.366 & 9.482 & KAIT & Clear & 17.22 $\pm$ 0.05 \\
2017-08-12.369 & 57977.369 & 10.485 & KAIT & Clear & 17.22 $\pm$ 0.04 \\
2017-08-13.362 & 57978.362 & 11.478 & KAIT & Clear & 17.41 $\pm$ 0.05 \\
2017-08-14.362 & 57979.362 & 12.478 & KAIT & Clear & 17.48 $\pm$ 0.05 \\
2017-08-15.360 & 57980.360 & 13.476 & KAIT & Clear & 17.57 $\pm$ 0.04 \\
2017-08-16.355 & 57981.355 & 14.471 & KAIT & Clear & 17.66 $\pm$ 0.04 \\
2017-08-17.337 & 57982.337 & 15.453 & KAIT & Clear & 17.68 $\pm$ 0.09 \\
2017-08-18.347 & 57983.347 & 16.463 & KAIT & Clear & 17.82 $\pm$ 0.10 \\
2017-08-19.339 & 57984.339 & 17.455 & KAIT & Clear & 17.88 $\pm$ 0.05 \\
2017-08-20.334 & 57985.334 & 18.450 & KAIT & Clear & 17.95 $\pm$ 0.06 \\
2017-08-22.333 & 57987.333 & 20.449 & KAIT & Clear & 18.13 $\pm$ 0.07 \\
2017-08-23.335 & 57988.335 & 21.451 & KAIT & Clear & 18.19 $\pm$ 0.13 \\
2017-08-24.316 & 57989.316 & 22.432 & KAIT & Clear & 18.31 $\pm$ 0.07 \\
2017-08-25.336 & 57990.336 & 23.452 & KAIT & Clear & 18.33 $\pm$ 0.06 \\
2017-08-26.339 & 57991.339 & 24.455 & KAIT & Clear & 18.41 $\pm$ 0.09 \\
2017-08-27.307 & 57992.307 & 25.423 & KAIT & Clear & 18.42 $\pm$ 0.06 \\
2017-08-28.320 & 57993.320 & 26.436 & KAIT & Clear & 18.61 $\pm$ 0.06 \\
2017-08-29.291 & 57994.291 & 27.407 & KAIT & Clear & 18.74 $\pm$ 0.15 \\
2017-08-30.278 & 57995.278 & 28.394 & KAIT & Clear & 18.47 $\pm$ 0.22 \\
2017-08-31.284 & 57996.284 & 29.400 & KAIT & Clear & 18.62 $\pm$ 0.17 \\
\hline
\hline
\end{tabular}
\end{table*}

\begin{table*}
\caption{Photometry of AT\,2017fvz from ATLAS observations.$^\mathrm{a}$}
\centering
\label{Photometry3}
\begin{threeparttable}
\begin{tabular}{l c r c c r}
\hline
\hline
UT Date & MJD (d) & $t-t_0$ (d) & Telescope \& instrument & Filter & Photometry (mag) \\

\hline
\hline
2017-08-03.389 & 57968.389 & 1.505 $\pm$ 0.005 & ATLAS & Orange & 16.511 $\pm$ 0.095 \\
2017-08-04.440 & 57969.440 & 2.556 $\pm$ 0.006 & ATLAS & Orange & 16.219 $\pm$ 0.036 \\
2017-08-09.455 & 57974.455 & 7.571 $\pm$ 0.005 & ATLAS & Orange & 16.955 $\pm$ 0.068 \\
2017-08-11.436 & 57976.436 & 9.552 $\pm$ 0.005 & ATLAS & Orange & 17.315 $\pm$ 0.032 \\
2017-08-12.412 & 57977.412 & 10.528 $\pm$ 0.005 & ATLAS & Orange & 17.447 $\pm$ 0.023 \\
2017-08-13.446 & 57978.446 & 11.562 $\pm$ 0.005 & ATLAS & Orange & 17.502 $\pm$ 0.015 \\
2017-08-15.424 & 57980.424 & 13.540 $\pm$ 0.005 & ATLAS & Orange & 17.520 $\pm$ 0.031 \\
2017-08-18.399 & 57983.399 & 16.515 $\pm$ 0.005 & ATLAS & Orange & 17.958 $\pm$ 0.098 \\
2017-08-22.384 & 57987.384 & 20.500 $\pm$ 0.005 & ATLAS & Orange & 18.209 $\pm$ 0.241 \\
2017-08-23.405 & 57988.405 & 21.521 $\pm$ 0.005 & ATLAS & Orange & 18.544 $\pm$ 0.319 \\
2017-08-26.394 & 57991.394 & 24.510 $\pm$ 0.004 & ATLAS & Orange & 20.143 $\pm$ 0.239 \\
2017-08-28.370 & 57993.370 & 26.486 $\pm$ 0.004 & ATLAS & Orange & 19.877 $\pm$ 0.118 \\
\hline
2017-08-16.394 & 57981.394 & 14.510 $\pm$ 0.005 & ATLAS & Cyan & 18.750 $\pm$ 0.169 \\
2017-08-17.417 & 57982.417 & 15.533 $\pm$ 0.005 & ATLAS & Cyan & 18.519 $\pm$ 0.442 \\
2017-08-21.411 & 57986.411 & 19.527 $\pm$ 0.010 & ATLAS & Cyan & 18.134 $\pm$ 0.410 \\
2017-09-17.329 & 58013.329 & 46.445 $\pm$ 0.004 & ATLAS & Cyan & 20.376 $\pm$ 0.403 \\
\hline
\hline
\end{tabular}
\begin{tablenotes}
\small
\item $^\mathrm{a}$The `orange' filter covers the $r'$ and $i'$ bands. The `cyan' filter covers the $V$ and $r'$ bands.
\item $^\mathrm{b}$The date listed here is the mean time of multiple observations taken on this date.
\item $^\mathrm{c}$The magnitude listed here is the mean magnitude calculated from multiple observations taken on this date with the associated standard uncertainty.
\end{tablenotes}
\end{threeparttable}
\end{table*}

\bsp
\label{lastpage}
\end{document}